\documentclass[12pt]{article}
\pdfoutput=1

\usepackage{graphicx}
\usepackage{float}
\usepackage{cite}

\usepackage{amsmath,bm}
\usepackage{amssymb}
\usepackage{amsfonts}         

\usepackage[bookmarks=true,colorlinks,linkcolor=red,urlcolor=blue,citecolor=blue]{hyperref}
\usepackage{setspace}

\renewcommand{\baselinestretch}{1.3}

\newcommand{\ie}{{\it i.e.}}

\newcommand{\onematrix}{1\!\!1}
\newcommand{\dbar}{\hspace{2pt}\mathrm{d}\hspace{-11pt}-\hspace{-3pt}}
\newcommand{\sign}{\,\mathrm{sign}\,}
\newcommand{\Tr}{\,\mathrm{Tr}\,}
\newcommand{\expval}[1]{\left\langle #1 \right \rangle}
\newlength{\slashoffset}
\newcommand{\ovslash}[1]{%
\settowidth{\slashoffset}{$#1\slash$}%
\addtolength{\slashoffset}{-1pt}%
#1\hspace{-0.5\slashoffset}\slash%
\hspace{0.5\slashoffset}%
\hspace{-5pt}%
}
\newcommand{\Dslash}{\ovslash{D}}
\newcommand{\norm}[1]{\left| #1\right|}
\newcommand{\bra}[1]{\left\langle #1\right|}
\newcommand{\ket}[1]{\left| #1\right\rangle}
\newcommand{\braket}[2]{\left\langle #1 | #2 \right\rangle}
\newcommand{\ii}{\mathrm{i}}
\newcommand{\dd}{\, \mathrm{d}}
\renewcommand{\(}{\left(}
\renewcommand{\)}{\right)}
\renewcommand{\[}{\left[}
\renewcommand{\]}{\right]}

\begin{document}

\begin{titlepage}

\begin{flushright}
MIT-CTP/4469; 
UCSD-PTH-13-08
\end{flushright}
\vfil

\begin{center}
{
\huge
{\bf 
How to construct \\
\vskip.1in
a gravitating quantum electron star}}\\
\end{center}
\vfil
\begin{center}
{\large Andrea Allais${}^a$ and 
John McGreevy${}^{b}$\footnote
{On 
leave from: Department of Physics, MIT,
Cambridge, Massachusetts 02139, USA.}}\\
\vspace{2mm}
${}^{a}$ Center for Theoretical Physics, MIT,
Cambridge, Massachusetts 02139, USA\\
${}^{b}$ Department of Physics, 
University of California at San Diego,
La Jolla, CA 92093, USA.
\vspace{4mm}
\end{center}

\vfil

\begin{center}

{\large Abstract}
\end{center}

\noindent
Motivated by the holographic study of Fermi surfaces, 
we develop methods to solve
Einstein gravity coupled to 
fermions and gauge fields,
with AdS boundary conditions
and a chemical potential.
\vfill
\begin{flushleft}
June 2013
\end{flushleft}
\vfil
\end{titlepage}
\newpage
\renewcommand{\baselinestretch}{1.1}  

\renewcommand{\arraystretch}{1.5}

\tableofcontents

\vfill\eject
\section{Introduction}

In this paper we address the following question:
what asymptotically anti de Sitter (AdS) spacetimes result from a 
finite density of gravitating, charged fermions?
While this question is a natural one in the context of the study of 
gravity solutions with the covariant infrared cutoff provided by AdS, there is also a second set of motivations coming from condensed matter physics. 

There is considerable experimental evidence for the existence of materials whose electronic structure cannot be described by Fermi liquid theory, nor by any other known effective theory. A signature of these exotic materials is the presence of a well defined Fermi surface, without long-lived quasiparticles. More precisely, as in ordinary Fermi liquids, the electron operator Green's function has a zero frequency singularity located over an entire surface in momentum space (the Fermi surface). However, contrary to Fermi liquid behavior, the relative frequency width of this resonance does not vanish as $k$ approaches the Fermi surface. 

The short life of the quasiparticles indicates that strong interactions are at play, a fact which stymies conventional theoretical investigation. The holographic duality \cite{Maldacena:1997re, Gubser:1998bc, Witten:1998qj}, on the other hand, provides a different starting point, far from weak copuling. Hence, the gravitational system we study in this paper, seen through the eye of the duality, could be a useful toy model displaying non Fermi liquid phenomenology. In the present scarcity of viable approaches, such a model would be very valuable even if it has exotic short-distance physics. 

\subsection{Statement of the problem}

Our target field theory is a relativistic CFT with a gravity dual,
a global U$(1)$ symmetry (our proxy for fermion number), 
 and a fermion operator charged under the symmetry (our proxy for bare electrons).
As a gravity dual, we are led to study asymptotically-AdS boundary conditions
on gravity coupled to quantum electrodynamics. We want to study the CFT at non-zero U$(1)$ charge density, so  we turn on a chemical potential $\mu$ for the U$(1)$ symmetry; this is encoded in the boundary conditions $ A_t|_\text{bound.} = \mu$.

Wilsonian naturalness suggests that we use the following action in the bulk (for a review
of the duality from this point of view, see {\it e.g.}~\cite{McGreevy:2009xe, Hartnoll:2011fn}):
\begin{equation}
\label{eq:bulkaction}
 Z = \int \[\mathcal D\(g, A, \psi\)\] \exp\[\ii \int\!\dd^4 x\, \sqrt{g} \(\frac{R-2\Lambda}{\kappa^2} - \frac{F^2}{4 q^2} + \bar \psi(\ii \Dslash - m)\psi \)\]\,
\end{equation} 
where $D$ is a derivative covariant under coordinate transformations
and U$(1)$ gauge transformations. We normalize the gauge field so that the fermion field has unit charge.

This model has been has been investigated in \cite{Lee:2008xf, NFL_from_holography, Cubrovic:2009ye, Faulkner:2009wj, electron_star, 
Cubrovic:2010bf, 
Puletti:2010de, Hartnoll:2010ik, Hartnoll:2010xj, 
Sachdev:2011ze, quantum_electron_star, 
Blake:2012tp,
Medvedyeva:2013rpa} under various (drastic) simplifying assumptions, which we briefly summarize:

The large $N$ limit in the CFT implies $\kappa^2 \Lambda \ll 1$ and $q \ll 1$, and suppresses the fluctuations of the metric and of the gauge field. 
The papers 
\cite{Lee:2008xf, NFL_from_holography,  
Cubrovic:2009ye, Faulkner:2009wj} work in the limit $\kappa \to 0$, $q \to 0$ with finite $\kappa / q$, where the backreaction of the fermions on the background can be ignored. This results in the Reissner-N\"ordstrom extremal black hole groundstate \cite{Romans:1991nq, Chamblin:1999tk}, with its $AdS_2$ near-horizon region and associated zero-temperature entropy.

In this approximation, one finds \cite{Faulkner:2009wj} fermion Green's functions of non Fermi liquid character, of the form
\begin{equation}
\label{eq:nflG} G \sim \frac{ 1}{ \omega^{2\nu} + |k|- k_F}\,.
\end{equation} 
However, the fact that the backreaction can be ignored means that the charge and energy density carried by the fermions is negligible compared to the charge and mass of the black hole. From the dual point of view, this describes a system in which the fermions that form a non-Fermi liquid are only a small fraction of all the degrees of freedom. The large bath 
which makes up most of the system has some issues: the finite zero-temperature entropy and associated instabilities encourage us to lift the too-strong assumption that bulk matter fields do not affect the dynamics.

In \cite{electron_star}, the bulk fermions are treated as a charged, gravitating fluid, in a Thomas-Fermi approximation. According to \cite{electron_star}, this approximation is valid in the limit of large fermion mass $m^2/\Lambda >> 1$, and 
results in a much less exotic IR geometry. However, a fermion with a large mass is dual to an operator with a large anomalous dimension, so this large mass limit leads to a somewhat unphysical class of field theories. Moreover, the associated large quantum numbers mean parametrically many Fermi surfaces in the fermion Green's function \cite{Hartnoll:2010xj}.

The paper \cite{Blake:2012tp} studies the 
limit of a large external magnetic field, where the bulk fermions 
are effectively one-dimensional and may be treated using bosonization.

\subsection{This paper} 

In this work, we will retain the full quantum nature of the fermionic field, while treating the path integrals over the metric and the gauge field in the saddle point approximation\footnote{Neglect of metric and gauge fluctuations, in this context, is equivalent to the Hartree-Fock approximation.}, as is natural at large $N$. 
It is reasonable to hope that bulk fermions with $m^2/ \Lambda \sim 1$,
which must be treated quantum mechanically, realize a happy medium between
the too-exotic AdS$_2$ solution, which results from no fermions, and the classical electron star, which results from heavy bulk fermions.

The leading contribution to the partition function comes from the on-shell action, evaluated on the field configuration that solves the equations of motion
\begin{equation}\label{the problem}
\left\{
\begin{aligned}
&D_\mu F^{\mu \nu} = q^2 J^\nu\\
&G_{\mu\nu} + \Lambda g_{\mu \nu} = \kappa^2 \[T^{(\psi)}_{\mu\nu} + T_{\mu\nu}^{(A)}\]\,,
\end{aligned}
\right.
\end{equation} 
where
\begin{align}
&J^\mu = \expval{\bar \psi \gamma^\mu \psi}\\
&T^{(\psi)}_{\mu\nu} = \expval{\bar \psi \gamma_{(\mu}\ii D_{\nu)} \psi}\\
&T^{(A)}_{\mu\nu} = \frac{1}{q^2}\[F_{\mu\rho}F^{\rho}{}_{\nu} - \frac{1}{4} g_{\mu\nu} F_{\alpha\beta} F^{\beta\alpha}\]\,.
\end{align} 
The expectation values are computed with respect to the fermionic field path integral
\begin{equation}
 \expval{\,\cdots\,} = \frac{1}{Z_\psi} \int [\mathcal D \psi]\,\cdots\,\exp\[\ii \int\!\dd^4 x\, \sqrt{g}\, \bar \psi(\ii \ovslash{D} - m)\psi\]\,.
\end{equation}

The solution of the saddle point equations constitutes a rather difficult problem, because the fermion current and stress tensor are {\it non-local} functionals of the background fields $g$ and $A$, for which there is no hope of finding an explicit closed form expression. In this sense the system (\ref{the problem}) is more similar to an integro-differential system of equations than to a system of differential equations.

Like for other integro-differential systems, a solution can be found numerically via an iterative approach. Starting from some configuration of $g$ and $A$, one computes numerically the corresponding fermion current and stress tensor. Then one uses these to construct a new configuration of $g$ and $A$ via (\ref{the problem}), and repeats the process. Eventually, one hopes, the process converges to a fixed point, which is a solution of the system of equations. 

Once a solution to the field equations is found, small perturbations about it can deliver all the correlators of the dual field theory.
This approach was used successfully in the frozen-geometry approximation 
(valid when $\kappa \to 0$)
in \cite{Sachdev:2011ze, quantum_electron_star}.

We emphasize that this iterative scheme for solving the saddle point equations has two separate parts:
\begin{enumerate}
\item Given a fixed set of currents\footnote{For brevity, we refer collectively to the charge current and the stress tensor as currents.} $ \expval{j^\mu}, \expval{T^\mu{}_\nu}$, 
solve (\ref{the problem}) to
find the gauge field $A$ and geometry $g$. This is a relatively standard problem,
which we treat in \S\ref{sec:einstein}.
\item Given the background $A, g$, evaluate the currents.
This problem is more difficult and less familiar. We will lavish a great deal of attention on it,
in \S\ref{sec:currents}.
\end{enumerate}

In \S\ref{sec:results} we present our results: the first gravitating quantum electron stars without a large magnetic field. The eager reader may safely read this last section first, as reference to \S\ref{sec:currents} and \S\ref{sec:einstein} is kept to a minimum.

Further details can be found in appendices. Of particular note is the appendix \ref{sec:cautionary}, where we describe the many alluring ways in which one should {\it not} approach the construction a gravitating quantum electron star.

\section{Computation of the fermionic currents}

\label{sec:currents}

\subsection{Regularization and renormalization}

In the continuum, the currents are divergent quantities. They must be regulated and then renormalized while preserving the symmetries of the low energy theory, that is: gauge invariance and general covariance \cite{birrel_davies}. The simplest way of doing so, at least conceptually, is to use a covariant regulator. For example, one could set up the problem in euclidean space, and use a heat kernel regulator to define the bare currents:
\begin{align}
&J^\mu_0 = \expval{\bar \psi \gamma^\mu e^{-s^2 D\!\!\!\!/^2}\psi}\\
&T^{\mu\nu}_0 = \expval{\bar \psi \gamma^{(\mu}\ii D^{\nu)} e^{-s^2 D\!\!\!\!/^2} \psi}\,.
\end{align} 

With this choice of regulator, the bare currents have the following small-$s$ expansion
\begin{align}
&J^\mu_0 = c_1 \log \frac{s}{L_{\text{IR}}} D_\nu F^{\mu\nu} + J^\mu_R + \mathcal{O}(s^2)\\
\begin{split}
&T^{\mu\nu}_0 =  \frac{c_2}{s^4} g^{\mu\nu} + \frac{1}{s^2} \(c_3 G^{\mu\nu} + c_4 m^2 g^{\mu\nu}\)\\ 
&\phantom{T^{\mu\nu}_0 = } + \log \frac{s}{L_{\text{IR}}}\(c_5\, H^{(1)\mu\nu} + c_6\, H^{(2)\mu\nu} + c_7\, m^2 G^{\mu\nu} + c_8\, m^4 g^{\mu\nu} + c_9\, q^2 T^{(A)\mu\nu}\)\\
&\phantom{T^{\mu\nu}_0 = } +T^{\mu\nu}_R + \mathcal{O}(s^2)\,,
\end{split}
\end{align} 
where the coefficients $c_i$ are (known) rational multiples of $1/\pi^2$, $H^{(1)}$ and $H^{(2)}$ are tensors involving four derivatives of the metric, and $T^{(A)}$ is the Maxwell stress tensor. $L_{\text{IR}}$ is some infrared renormalization scale of choice: changing it amounts to a finite renormalization of coupling constants, as described below.

The divergent terms in the series are local functionals of the background. This is because they come from high energy, short wavelength modes, which are sensitive only to the local physics. They are geometric objects with all the necessary symmetries: they are covariantly conserved tensors, with the right dimension, and transform appropriately under charge conjugation. This is because the regulator is gauge- and diffeomorphism-invariant.

By looking at (\ref{the problem}), it is clear that the divergent terms can be absorbed in the renormalization of: the charge $q$, the cosmological constant $\Lambda$ and Newton's constant $\kappa^2$, with the exception of $H^{(1)}$ and $H^{(2)}$. These renormalize two higher derivative corrections to the Einstein-Hilbert action. In fact
\begin{align}
 & H^{(1)\mu\nu}= \frac{1}{\sqrt{g}}\frac{\delta}{\delta g_{\mu\nu}} \int \dd^4 x\, \sqrt{g}\, R^2\,, \\
 & H^{(2)\mu\nu}= \frac{1}{\sqrt{g}}\frac{\delta}{\delta g_{\mu\nu}} \int \dd^4 x\, \sqrt{g}\, R^{\alpha\beta}R_{\alpha\beta}\,.
\end{align} 
In the spirit of retaining only the very low energy physics of the theory, we will set the renormalized coefficient of these higher derivative terms to zero.

After appropriate renormalization of the couplings, we are left with the renormalized, finite currents $J_R$ and $T_R$, which is what we take to stand on the right hand side of (\ref{the problem}). These quantities receive contributions from the whole spectrum of modes of the fermionic field, they are sensitive to the infrared physics, and therefore they are non-local, non-geometric functionals of the background.

Although the regularization and renormalization prescription described above is conceptually very simple, it proved unfeasible to follow in practice. For various technical reasons, the heat kernel regularization, or any other covariant method like dimensional, zeta function or Pauli-Villars regularization, turns out not to be well suited for the numerical computation we require (see section \ref{other regulators}). 

Instead, we resorted to point splitting regularization. 
Starting from a point $x$, we shoot out a geodesic, in a direction specified by a unit vector $t$, and we take a point $x'$ along it, at a geodesic distance $s$ from $x$. We then define the regularized current and stress tensor at $x$ as
\begin{align}
&J^\mu_0(x) = \expval{\bar \psi(x') \gamma^\mu \psi(x)}\\
&T^{\mu\nu}_0(x) = \expval{\bar \psi(x') \gamma^{(\mu}\ii D^{\nu)} \psi(x)}\,.
\end{align} 

A small-$s$ expansion has been worked out for these quantities also. However, because the regulator breaks gauge and diffeomorphism invariance, it involves contractions of local geometric tensors with the vector $t$, and the terms are not covariantly conserved. In the massless case $m = 0$ it has the form\footnote{The precise form of the expansion depends on the details of the point splitting prescription. This formula is taken from \cite{christensen}, and is derived according to the definitions there.}
\begin{equation}\label{adiabatic expansion}
\begin{split}
 T_0^{\mu\nu} = & -\frac{1}{\pi^2 s^4}\(g^{\mu\nu} - 4 t^\mu t^\nu\)\\
 &-\frac{1}{16 \pi^2 s^2} \[\frac{2}{3}\(G^{\mu\nu} + R\, t^\mu t^\nu\) - 4 R^{(\mu}{}_{\lambda}\, t^{\nu)} t^{\lambda} + \frac{4}{3} \(g^{\mu\nu} R_{\lambda \xi}- R^\mu{}_\lambda{}^\nu{}_\xi\)\, t^\lambda t^\xi\]\\
 &-\frac{1}{160 \pi^2}\log\frac{s}{L_{\text{IR}}}\(H^{(2)\mu\nu} - \frac{1}{3}H^{(1)\mu\nu}\) + T_{\text{finite}}^{\mu\nu}(t) + \mathcal{O}(s^2)\,.
\end{split}
\end{equation} 

It is very important to stress that $T_{\text{finite}}$, besides being a non-local, non-geometric object, still depends on the vector $t$, {\it i.e.} on the regularization scheme, and hence cannot be interpreted as a renormalized quantity. To obtain a well defined renormalized stress tensor, it is necessary to proceed with what is called adiabatic renormalization \cite{birrel_davies}.

Adiabatic renormalization consists in subtracting from $T_{\text{finite}}$ an additional, finite counterterm, that precisely compensates for all the symmetry breaking effects of the regulator. To determine what this counterterm is, it is necessary to compute the bare stress tensor within a derivative, or adiabatic, expansion: the background is assumed to change on length scales much larger than the correlation length of the fermion field, set by the mass $m$. 

Within this approximation, it is possible to compute all the orders of the expansion (\ref{adiabatic expansion}), including $T_\text{finite}$, explicitly as functionals of the background. They all turn out to be local terms, made of contractions of geometric tensors with the vector $t$. This means that the adiabatic expansion completely misses the non-local, infrared physics. However, it retains all the symmetry breaking effects of the regulator, which affects only local, UV physics. Therefore, by subtracting from the bare stress tensor $T_0$ its adiabatic expansion, up to and including the finite term, and taking the limit $s \to 0$, a well defined renormalized stress tensor $T_R$ is obtained. When other covariant regularization procedures are feasible, it has been shown that this approach yields the same renormalized stress tensor $T_R$, up to a finite renormalization of $\Lambda$, $\kappa^2$ and the other couplings. The current can be regularized and renormalized according to the same procedure.

\subsection{The conformal anomaly and covariant conservation}
\label{conservation section}

An important check of the regularization and renormalization prescription we are adopting is the ability to produce a covariantly conserved stress tensor. In a static geometry the bare stress tensor, defined by
\begin{equation}
 T^{\mu\nu}_0(x) = \expval{\bar \psi(x') \gamma^{(\mu}\ii D^{\nu)} \psi(x)}\,,
\end{equation} 
is covariantly conserved, if $x - x' = (\Delta t, 0, 0, 0)$ with constant $\Delta t$. In fact we have
\begin{equation}
\begin{split}
 D_\mu T^{\mu \nu} &= \partial_\mu T^{\mu \nu} + \Gamma^{\mu}{}_{\mu \rho} T^{\rho\nu} + \Gamma^{\nu}{}_{\mu \rho} T^{\mu\rho}\\
&=\expval{\bar \psi(x') \overrightarrow{\partial}_\mu \gamma^{(\mu}\ii D^{\nu)} \psi(x)} + \expval{\bar \psi(x') \overleftarrow{\partial}_\mu \gamma^{(\mu}\ii D^{\nu)} \psi(x)} +\\ 
&\phantom{=} +\Gamma^{\mu}{}_{\mu \rho} T^{\rho\nu} + \Gamma^{\nu}{}_{\mu \rho} T^{\mu\rho}\,,
\end{split}
\end{equation}
and, because the sections do not depend on time, the derivatives can be promoted to covariant derivatives:
\begin{equation}
\begin{split}
 D_\mu T^{\mu \nu} &= \expval{\bar \psi(x') \overrightarrow{D}_\mu \gamma^{(\mu}\ii D^{\nu)} \psi(x)} + \expval{\bar \psi(x') \overleftarrow{D}_\mu \gamma^{(\mu}\ii D^{\nu)} \psi(x)}\,.
\end{split}
\end{equation}
Using the field equations of motion, it is easy to show that this quantity is zero. It is also possible to show directly that the stress tensor computed with the adiabatic expansion is covariantly conserved, and so is the renormalized stress tensor.

If another point-splitting prescription is taken, for example one in which $\Delta t$ depends on position, then neither the bare stress tensor, nor the adiabatic stress tensor are conserved. However, the difference of the two, in the limit $s \to 0$ is conserved.

Another important check is the manifestation of the conformal anomaly. In fact, it is well known  that, for a massless field, the trace of the stress tensor on curved spacetime is not zero, but is proportional to a local geometric functional of the background. Using the equations of motion, it is easy to show that
\begin{equation}
 T^\mu{}_\mu(x, x') = m\, \expval{\bar\psi(x) \psi(x)} + \text{contact terms}\,,
\end{equation} 
and it would be natural to conclude that, in the massless limit, the trace is zero. This is certainly true for the bare stress tensor. On the other hand, the adiabatic expansion assumes the correlation length $1/m$ to be much smaller than the other length scales, and hence it breaks down in the massless limit. This becomes manifest as a $1/m$ divergence in $\expval{\bar\psi(x) \psi(x)}$, which cancels the factor of $m$, and gives a finite contribution in the massless limit. Since the renormalized stress tensor is the difference between the bare and the adiabatic quantity, it also acquires a finite trace in the massless limit. The trace of the stress tensor obtained in this way correctly reproduces the conformal anomaly
\begin{equation}
 T^\lambda{}_\lambda = \frac{1}{2880 \pi^2} \( -\frac{7}{4} R_{\mu\nu\rho\sigma}R^{\mu\nu\rho\sigma} - 2 R_{\mu\nu}R^{\mu \nu} + \frac{5}{4} R^2 - 3 \Box R\)\,.
\end{equation} 

\subsection{A choice of background}

It is not possible to carry out the computation of the bare currents in a completely arbitrary background, even with time-translation invariance and spatial rotation invariance.
Even restricting to an asymptotically anti de Sitter space is not enough: some knowledge of the the interior is needed. Therefore, we choose to target a class of metrics
\begin{equation}\label{the background}
 g = -e_t^2(r) \dd t^2 + e_r^2(r) \dd r^2 + e_s^2(r) \dd \Omega_2^2\,,
\end{equation} 
that are smoothly connected to global anti de Sitter space
\begin{equation}
 g = \frac{L^2}{\frac{4}{\pi^2}\cos^2 \frac{\pi r}{2}} 
\(-\dd t^2 + \dd r^2 + \frac{4}{\pi^2}\sin^2 \frac{\pi r}{2} \dd \Omega_2^2\)\,.
\end{equation} 
That is, we take $r \in (0, 1)$, with $e_s$ vanishing linearly at $r = 0$ and all the sections $e_\mu$ diverging like $(1 - r)^{-1}$ at $r = 1$. It is useful to think of the spatial sections of this class of metrics as 3-balls, with the center at $r = 0$, and the edge at $r = 1$, where the conformal factor diverges. 

The spherical spatial sections should be regarded as a (covariant and natural) IR regulator,
which replaces the artificial hard wall $z_m$ employed in 
\cite{quantum_electron_star}, following \cite{Sachdev:2011ze}.
Such a regulator has been used to good effect in the analogous problem
of charged scalar fields in AdS \cite{Gentle:2011kv}\footnote
{We thank Simon Gentle for useful discussions of this point.
}.
A non-covariant regulator such as a hard wall is an obstruction to  
building a covariant bulk stress tensor.
This regulator has the further virtue (in contrast to the hard wall) of uniquely specifying the IR boundary conditions
on the bulk spinor fields, simply by regularity.

A further practical reason for this choice is that this class of spaces is compact from the point of view of the Dirac Hamiltonian, which therefore has a discrete spectrum and normalizable eigenfunctions. This is a big advantage for a numerical computation, which is lost, for example, in spaces with a horizon in the interior.

We will take the gauge field to have the form
\begin{equation}
 A = \Phi(r) \dd t\,,
\end{equation} 
and the chemical potential $\mu$ sets the boundary condition for $\Phi$:
\begin{equation}
 \Phi(1) = \mu\,.
\end{equation} 

From the dual point of view, this choice of background is analogous to defining the field theory on a sphere of radius $R = \alpha(1)L$, instead of flat space. The sphere is just an infrared regulator, whose effect becomes negligible in the limit $\mu \gg \frac{1}{R}$.

The fermion field is a free field, so all the information is contained in the Green's function
\begin{equation}
 S(x, x') \equiv \expval{\psi(x) \bar \psi(x')}\,,
\end{equation} 
which satisfies
\begin{equation}
 \sqrt{g}\,(\ii \gamma \cdot D - m) S(x, x') = \ii\, \delta(x - x')\,.
\end{equation} 
The currents, in terms of the Green's function, are given by
\begin{align}
&j^\mu(x) = - \Tr[\gamma^\mu S(x, x')]\,,\\
& T^{\mu\nu}(x) = - \Tr\[\gamma^{(\mu}\, \ii D^{\nu)} S(x, x')\]\,.
\end{align}

Specifying to our background, we introduce the Hamiltonian Dirac matrices
\begin{align}
 &\alpha^t = \gamma^t\,,
 &&\alpha^i = \gamma^t \gamma^i\,,
 &&(\alpha^\mu)^\dag = \alpha^\mu\,,
 && \{\alpha^\mu, \alpha^\nu\} = \delta^{\mu\nu}\,,
\end{align} 
and the covariant derivatives are
\begin{align}
 & D_t = \partial_t + \ii \Phi + \frac{e'_t}{2 e_r} \alpha^r\,,\\
 & D_r = \partial_r\,,\\
 & D_\theta = \partial_\theta - \frac{e'_s}{2 e_r} \alpha^r\alpha^\theta\,,\\
 & D_\phi = \partial_\phi - \sin\theta \frac{e'_s}{2 e_r} \alpha^r\alpha^\phi - \frac{1}{2} \cos \theta \alpha^\theta \alpha^\phi\,.
\end{align}

Since our background is static, it is advantageous to move to a Hamiltonian picture, by defining
\begin{equation}
 S(x, x') = \[g^{-\frac{1}{4}}\sqrt{e_t}\]_x G(x, x')\alpha^t \[g^{-\frac{1}{4}} \sqrt{e_t} \]_{x'}\,,
\end{equation} 
so that $G$ satisfies the equation
\begin{equation}
  \(\ii \partial_t - H \) G(x, x') = \ii\,\delta(x - x')\,,
\end{equation} 
with
\begin{align}
 H = -\ii \frac{e_t}{e_r} \alpha^r \[\partial_r + \frac{e'_t}{2e_t} -  \frac{e'_r}{2e_r}\]  - \ii \frac{e_t}{e_s} \[\alpha^\theta \partial_\theta + \alpha^\phi \frac{1}{\sin \theta} \partial_\phi \] + \Phi + m e_t \alpha^t\,.
\end{align} 

The Hamiltonian $H$ is self-adjoint with respect to the scalar product
\begin{equation}
 (\psi_1, \psi_2) = \int \dd r \dd \theta \dd\phi\ \psi_1^\dag(r, \theta, \phi) \psi_2(r, \theta, \phi)\,.
\end{equation} 

In our background, only the time component of the current and the diagonal components of the stress tensor have a non-zero expectation value. In terms of the Hamiltonian Green's function $G$ these are given by
\begin{align}
& J^t = - f_{x,x'} \Tr\[G(x, x')\]\,,\\
& T^t{}_t = - \ii\, f_{x,x'}\Tr\[\(\partial_t + \ii \Phi + \frac{e'_t}{2e_r} \alpha^r\)G(x, x')\]\,,\\
& T^r{}_r = - \ii\, f_{x,x'}\frac{e_t}{e_r}\Tr\[\(\alpha^r \partial_r - \alpha^r \frac{e'_r}{2e_r} - \alpha^r \frac{e'_s}{e_s}\)G(x, x')\]\,,\\
& T^\theta{}_\theta = - \ii\, f_{x,x'}\frac{e_t}{e_s}\Tr\[\(\alpha^\theta \partial_\theta -\alpha^\theta \frac{\cos\theta}{2\sin\theta} + \alpha^r \frac{e'_s}{2e_r} \)G(x, x')\]\,,\\
& T^\phi{}_\phi = - \ii\, f_{x,x'}\frac{e_t}{e_s}\Tr\[\(\alpha^\phi \frac{\partial_\phi}{\sin\theta} +\alpha^\theta \frac{\cos\theta}{2\sin\theta} + \alpha^r \frac{e'_s}{2e_r} \)G(x, x')\]\,,
\end{align}
where
\begin{equation}
 f_{x,x'} = g^{-\frac{1}{4}}(x) g^{-\frac{1}{4}}(x') \frac{e_t(x')}{e_t(x)}\,,
\end{equation} 
and all other sections are evaluated at $x$.

\subsection{Adiabatic expansion}

We now show how the Green's function, and hence the bare currents, can be computed within a small derivative, or adiabatic, expansion \cite{bunch}. 

To outline the idea behind the computation, it is useful to consider the simpler problem
\begin{equation}\label{simple problem}
 \[-\nabla^2 + m^2 + V(x)\] G(x, x') = \delta(x - x')\,.
\end{equation} 
We take $V$ to be varying slowly compared to the correlation length $1/m$, and we  expand $V(x)$ about $x'$. Then, equation (\ref{simple problem}) can be written symbolically as
\begin{equation}
 \[G_0^{-1} + A\] G = \onematrix,
\end{equation} 
with
\begin{align}
 &G_0^{-1} = -\nabla^2 + m^2 \,, \\
 &A = V_{,i}(x')(x - x')_i + \frac{1}{2} V_{,ij}(x')(x - x')_i (x - x')_j + \ldots\,,
\end{align} 
and it is formally solved by a series in $A$:
\begin{equation}
 G = G_0 \sum_{n = 0}^{\infty} (- A G_0)^n\,.
\end{equation} 

The matrix products in this series actually stand for convolutions, so it is useful to go to momentum space:
\begin{equation}
 G(x; x') = \int \dbar^d k\ e^{\ii k \cdot (x - x')}G(k; x')
\end{equation} 
where $ \dbar k \equiv { \dd k \over 2\pi }$.  So
\begin{align}
 &G_0^{-1} = k^2 + m^2\,, \\
 &A = V_{,i}(x')\, \ii \partial_{k_i} - \frac{1}{2} V_{,ij}(x') \partial_{k_i} \partial_{k_j} + \ldots\,.
\end{align} 

Using the identity
\begin{align}
 &\partial_{k_i} G_0 = -2 k_i G_0^2\,,
\end{align}
we have
\begin{equation}
 G(k; x') = G_0 + (2\ii V_{,i} k_i - V_{,ii}) G_0^3 + (4V_{,ij}k_ik_j+2V_{,i}V_{,j}) G_0^4  -12 V_{,i}V_{,j} k_i k_j G_0^5+\ldots\,,
\end{equation} 
where we have retained terms involving up to two derivatives of the potential. Now we revert to position space. We have
\begin{align}
 \int \dbar^d k\ e^{\ii k \cdot s }G_0(k) = \frac{1}{(2\pi)^{\frac{d}{2}}} F_{d-2}(s, m)\,,&&
 F_{n}(s, m) = \(\frac{m}{s}\)^{\frac{n}{2}} K_{\frac{n}{2}}(m s)\,,
\end{align} 
where $s_i = x_i - x'_i$ and $s = \sqrt{s_i s_i}$, and we use the identities
\begin{align}
 &\int \dbar^d k\ e^{\ii k \cdot s }G_0^n(k) = -\frac{1}{2m(n-1)} \frac{\partial}{\partial m} \int \dbar^d k\ e^{\ii k \cdot s }G_0^{n - 1}(k)\,,\\
 &\int \dbar^d k\ k_i e^{\ii k \cdot s }f(k) = -\ii \partial_{s_i} \int \dbar^d k\ e^{\ii k \cdot s }f(k)
\end{align} 
to carry out the fourier transform. The final result is
\begin{equation}
\begin{split}
 (2\pi)^{\frac{d}{2}} G(x, x') =& F_{d-2} -\(\frac{1}{4} V_{,i} s_i +\frac{1}{12} V_{,ij} s_i s_j\) F_{d-4} -\\
 & - \(\frac{1}{24} V_{,ii} - \frac{1}{32} V_{,i} V_{,j} s_i s_j \) F_{d-6}  + \frac{1}{96} V_{,i}V_{,i} F_{d-8} + \ldots
\end{split}
\end{equation} 
Then we can expand in series for small $s$, and we have, for $d = 4$
\begin{equation}\label{sample adiabatic expansion}
\begin{split}
 (2\pi)^{2} G(x, x') & = \frac{1}{s^2} + \frac{m^2}{4} L - \frac{m^2}{4} -\frac{1}{24} V_{,ii}{m^2} + \frac{1}{48} \frac{V_{,i}V_{,i}}{m^4} +\\
 &+ \frac{1}{8} V_{,i} s_i L + \frac{m^4}{32} s^2 L - \frac{1}{96} V_{,ii} s^2 L + \frac{1}{24} V_{,ij} s_i s_j L + \mathcal O(s^2)\,,
\end{split}
\end{equation} 
where $L = \log m^2 s^2/4 +\gamma_E$. A structure similar to that of eq. (\ref{adiabatic expansion}) starts to be apparent. 

A computation along the same lines can be carried out for the fermionic Green's function
\begin{equation}
  \(\ii \partial_t +\ii \frac{e_t}{e_r} \alpha^r \[\partial_r + \frac{e'_t}{2e_t} -  \frac{e'_r}{2e_r}\]  + \ii \frac{e_t}{e_s} \[\alpha^\theta \partial_\theta + \alpha^\phi \frac{1}{\sin \theta} \partial_\phi \] - \Phi - m e_t \alpha^t \) G = \ii\,\delta\,.
\end{equation} 
It involves the same steps, including the integrals (after Wick rotation), but is algebraically much messier. In fact, it is necessary use a computer algebra system, and we found it advantageous to specify the direction of point splitting from the beginning. Once the Green's function is found, the currents can be computed by taking the opportune derivatives. The result is 
reported in appendix \ref{expansion appendix}.

In alternative to the method described, there are also covariant ways to carry out the adiabatic expansion \cite{christensen}, which, however, are more difficult to implement on the computer.

\subsection{Computation of the bare currents}

We compute the bare currents by expanding the Green's function on a basis of eigenfunctions of the Dirac Hamiltonian $H$. The problem can be reduced to one dimension by exploiting translational invariance in the time direction, and the spherical symmetry of the spatial sections. 

In order to exploit the
spherical symmetry, we must introduce spinor spherical harmonics. For the two-sphere, they are solutions of the eigenvalue equation
\begin{equation}
 \[\sigma^2(-\ii \partial_\theta) +  \sigma^1 \frac{1}{\sin \theta} (- \ii\partial_\phi)\] Y_{\ell m}(\theta, \phi) = \ell\, Y_{\ell m}(\theta, \phi)\,.
\end{equation} 
The spectrum is quantized, with $\ell \in \{+1,\, -1,\, +2,\, -2,\,\ldots\}$, and $m$ labels the degeneracy $2|\ell|$ of each eigenspace. The spinor harmonics are orthonormal and complete:
\begin{align}
 &\int\! \dd\theta \dd \phi\  Y^\dag_{\ell m}(\theta, \phi)Y_{\ell' m'}(\theta, \phi) = \delta_{\ell\ell'}\delta_{mm'}\,,\\
 &\sum_{\ell m}Y_{\ell m}(\theta, \phi)Y^\dag_{\ell m}(\theta', \phi') = \delta(\theta - \theta')\delta(\phi - \phi')\,,
\end{align}
and they also satisfy (note that the sum runs only over the degeneracy index $m$):
\begin{align}
 &\sum_{m} Y^\dag_{\ell m}(\theta, \phi)Y_{\ell m}(\theta, \phi) = \frac{\ell}{2\pi} \sin(\theta)\,,\\
 &\sum_{m} Y^\dag_{\ell m}(\theta, \phi)\sigma^2(-\ii \partial_\theta) Y_{\ell m}(\theta, \phi) = \sign \ell\frac{\ell^2}{4\pi} \sin(\theta)\,,\\
 &\sum_{m} Y^\dag_{\ell m}(\theta, \phi)\sigma^1\frac{-\ii \partial_\phi}{\sin \theta} Y_{\ell m}(\theta, \phi) = \sign\ell \frac{\ell^2}{4\pi} \sin(\theta) \,.
\end{align} 

We organize the Dirac matrices as follows:
\begin{align}
 &\alpha^r = \onematrix\otimes \sigma^2\,, 
 &&\alpha^\theta = \sigma^2 \otimes \sigma^3\,,
 &&\alpha^\phi = \sigma^1 \otimes \sigma^3\,,
 &&\alpha^t = \onematrix \otimes \sigma^1\,, 
\end{align}
and we exploit this direct product structure to write
\begin{equation}
 G(x, x') = \int \dbar \omega\ e^{- \ii \omega(t -t')}\sum_{\ell m} \[Y_{\ell m}(\theta, \phi) Y^\dag_{\ell m}(\theta', \phi')\] \otimes G_{\omega\ell}(r, r')\,,
\end{equation} 
where $G_{\omega \ell}$ is the Green's function of a simple one-dimension differential operator:
\begin{equation}
 \(\omega - H_{\ell} \) G_{\omega\ell}(r, r') = \ii \delta(r - r')\,,
\end{equation} 
with
\begin{align}
 H_{\ell} = -\ii \sigma^2 \frac{e_t}{e_r} \[\partial_r + \frac{e'_t}{2e_t} -  \frac{e'_r}{2e_r}\] + \ell\frac{e_t}{e_s} \sigma^3 + \Phi + m e_t \sigma^1\,.
\end{align} 

$H_\ell$ is a self-adjoint operator, with an orthonormal and complete set of real eigenfunctions. In the class of backgrounds we are considering, the spectrum is discrete, and we label it with an index $n$:
\begin{equation}
 H_\ell \psi_{n\ell}(r) = \omega_{n\ell} \psi_{n\ell}(r)\,.
\end{equation} 

Then we have
\begin{equation}
 G(x, x') = \int \dbar \omega\ \frac{\ii\, e^{- \ii \omega(t -t')}}{\omega - \omega_{n\ell}}\sum_{n \ell m} \[Y_{\ell m}(\theta, \phi) Y^\dag_{\ell m}(\theta', \phi')\] \otimes \[\psi_{n\ell}(r) \psi_{n\ell}^\dag(r')\]\,.
\end{equation} 

We perform Wick rotation, and we take the point-splitting to be along the imaginary time direction, i.e. $t - t' = \pm \ii s$. Then we symmetrize with respect to the sign of $s$ and we have
\begin{equation}
 G(x, x') = \sum_{n \ell m} \Theta_{n\ell}(s) \[Y_{\ell m}(\theta, \phi) Y^\dag_{\ell m}(\theta', \phi')\] \otimes \[\psi_{n\ell}(r) \psi_{n\ell}^\dag(r')\]\,,
\end{equation} 
where
\begin{equation}
 \Theta_{n\ell}(s) = \frac{1}{2} \sign \omega_{n\ell} e^{- |s \omega_{n\ell}|}\,.
\end{equation} 

Now we substitute in the expression for the currents. To keep things as easy as possible, we let $r = r'$, $\theta = \theta'$, $\phi = \phi'$, that is, we do not point-split in the
other directions. Using the spinor harmonics identities, and the reality of wavefunctions, we have
\begin{align}
J^t & = \frac{1}{e_t e_r e_s^2} \sum_{n \ell} \frac{|\ell|}{2\pi}\  \Theta_{n\ell}(s) \psi^\dag_{n\ell}(r)\psi_{n\ell}(r) \,,\\
T^t{}_t &= \frac{1}{e_t e_r e_s^2}\sum_{n \ell} \frac{|\ell|}{2\pi}\  \Theta_{n\ell}(s)  \(-\omega_{n\ell} + \Phi\)\psi^\dag_{n\ell}(r)\psi_{n\ell}(r)\,,\\
T^r{}_r& = \frac{1}{e_t e_r e_s^2} \sum_{n \ell} \frac{|\ell|}{2\pi}\ \Theta_{n\ell}(s) \frac{e_t}{e_r}\psi^\dag_{n\ell}(r)(-\ii \sigma^2)\psi'_{n\ell}(r)\,,\\
T^\theta{}_\theta = 
T^\phi{}_\phi &= \frac{1}{e_t e_r e_s^2}\sum_{n \ell} \frac{|\ell|}{2\pi}\ \Theta_{n\ell}(s) \frac{\ell}{2} \frac{e_t}{e_s} \psi^\dag_{n\ell}(r)\sigma^3\psi_{n\ell}(r)\,.
\end{align} 

With these manipulations, the computation of the bare currents has been reduced to the problem of diagonalizing the one-dimensional Hamiltonian $H_\ell$ and carrying out the mode sums above. Both tasks can be carried out numerically, and the second is feasible especially thanks to the exponential suppression of the high energy modes, due to the factor $\Theta_{n\ell}(s)$.

\subsection{Diagonalization of the Dirac Hamiltonian}

Let us now show how the Hamiltonian $H_\ell$ can be diagonalized numerically. The spectrum and the eigenfunctions must be computed with very high accuracy, and efficiently. Using a finite-differences discretization of the Hamiltonian is not sufficient for the purpose. It is necessary to resort to spectral methods \cite{spectral_methods}, which consist in approximating the eigenfunctions with polynomials of high degree, instead of a set of values on a uniformly-spaced grid. This allows a better representation of the derivative operator, and yields spectrum and eigenfunctions accurate to order $e^{-n}$, where $n$ is both the degree of the approximating polynomial and the rank of the matrix to be numerically diagonalized. This should be compared with the accuracy of finite-differences methods, which is only polynomial in $n$.

In order to use spectral methods, the metric must be further specified. In fact, the background (\ref{the background}) possesses residual reparametrization invariance, which we use to impose the constraint $e_t(r) = e_r(r)$, so that the metric takes the form
\begin{align}
 & g = \frac{1}{\beta^2(r)}\( -\dd t^2 + \dd r^2 + \alpha^2(r) \dd \Omega_2^2 \)\,,  && A = \Phi(r) \dd t\,.
\end{align}

Demanding a space with the same asymptotics as global AdS, we take
\begin{align}
 &\alpha \sim r\,, && \beta \sim b_0 r^0\,, && \text{for $r\to0$}\,,\\
 &\alpha \sim a_0 (1-r)^0\,, && \beta \sim \frac{1-r}{L}\,, &&\text{for $r\to1$}\,,\\
 &\alpha(-r) = -\alpha(r)\,, && \beta(-r) = \beta(r)\,, && \Phi(-r) = \Phi(r) \,.
\end{align}

The reparametrization invariance could be used to impose a different constraint instead of $e_r = e_t$, leading to different asymptotics of the sections. However, this choice has the big advantage that all the terms in the Hamiltonian
\begin{align}
 H_\ell = -\ii \sigma^2 \partial_r + \frac{\ell}{\alpha(r)} \sigma^1 + \Phi(r) + \frac{m}{\beta(r)} \sigma^3\,,
\end{align} 
are analytic for $r\in[0,1]$. This is crucial for the possibility of using spectral methods.

The eigenfunctions, however are not analytic at $r = 1$. To have a good polynomial approximation, the non-analytic behavior at the boundaries must be determined and factored out. For $r \to 0$, retaining only the leading terms, we have
\begin{align}
 & H \sim -\ii \sigma^2 \partial_r  + \frac{\ell}{r} \sigma^1\,, &&
 \psi(r) \sim \begin{pmatrix} a_1\\a_2 \end{pmatrix} r^\lambda\,,
\end{align} 
and there are two solutions:
\begin{align}
 a_1 = 0\,,\, \lambda = \ell\,,
 && \text{or} &&
 a_2 = 0\,,\, \lambda = -\ell\,.
\end{align}
The first solution is normalizable for $\ell > 0$, the second for $\ell < 0$. For $r \to 1$ we have
\begin{align}
 &H \sim -\ii \sigma^2 \partial_r  + \frac{mL}{1-r} \sigma^3\,,
&& \psi(r) \sim \begin{pmatrix} b_1\\b_2 \end{pmatrix} (1-r)^\nu\,.
\end{align} 
There are two solutions
\begin{align}
 b_1 = -b_2\,,\, \nu = m L\,,
 && \text{or} &&
 b_1 = b_2\,,\, \nu = -m L\,. 
\end{align}
The first solution is normalizable for $mL > - 1/2$, the second for $mL < 1/2$.

Finally, we notice that the Hamiltonian has parity symmetry:
\begin{equation}
 \sigma^3 H(-r) \sigma^3 = H(r)\,,
\end{equation} 
and hence the eigenfunctions can be taken to have definite parity
\begin{equation}\label{parity}
 \psi(-r) = \pm \sigma^3 \psi(r)\,.
\end{equation} 

We can collect all this information by writing the normalizable\footnote{Here we consider the case $mL > -\frac{1}{2}$. The other case $mL < +\frac{1}{2}$ is equivalent.} eigenfunctions as
\begin{equation}
 \psi(r) = r^{|\ell|} (1-r^2)^{mL} \phi(r)\,,
\end{equation} 
where $\phi(r)$ is analytic at $r = 0$ and at $r = 1$ and it satisfies
\begin{align}
 &\phi(-r) = - \sign(\ell) \sigma^3 \phi(r)\,,
 &\phi_1(1) + \phi_2(1) = 0\,.
\end{align}
Therefore, $\phi$ can be well approximated by a polynomial, and we can construct a complete basis 
of spinors with polynomial components
that satisfies the boundary conditions and parity constraints:
\begin{align}
 &\phi_a(r) = \begin{pmatrix} +Q_{a-1}(1) Q_a(r) \\ -Q_a(1) Q_{a-1}(r)\end{pmatrix} \quad \begin{aligned}&\text{odd $a$}\\& \ell > 0\end{aligned}&&
 \phi_a(r) = \begin{pmatrix}  -Q_a(1) Q_{a-1}(r) \\ +Q_{a-1}(1) Q_a(r)\end{pmatrix} \quad \begin{aligned}&\text{even $a$}\\&\ell > 0\end{aligned}\\
 &\phi_a(r) = \begin{pmatrix} -Q_a(1) Q_{a-1}(r) \\ +Q_{a-1}(1) Q_a(r)\end{pmatrix} \quad \begin{aligned}&\text{odd $a$}\\&\ell < 0\end{aligned}&&
 \phi_a(r) = \begin{pmatrix}  +Q_{a-1}(1) Q_a(r) \\ -Q_a(1) Q_{a-1}(r)\end{pmatrix} \quad \begin{aligned}&\text{even $a$}\\&\ell < 0\end{aligned}\,,
\end{align}
where the $Q_a$ are polynomials, such that $Q_a$ has degree $a \in \{0,1, 2,\ldots\}$ and the same parity as $a$. Since $H_\ell$ is self adjoint with respect to the scalar product
\begin{equation}
 \braket{\psi_1}{\psi_2} = \int_{-1}^{1} \dd r\ \psi_1^\dag(r) \psi_2(r) = \int_{-1}^{1} \dd r\ \ r^{2|\ell|} (1-r^2)^{2 m L} \phi_1^\dag(r) \phi_2(r)\equiv (\phi_1 , \phi_2)\,,
\end{equation} 
we take the polynomials $Q_a$ to be orthogonal with respect to the same scalar product:
\begin{equation}
 (Q_a, Q_b) = h_a \delta_{ab}\,.
\end{equation} 

These may be constructed from the Jacobi polynomials as follows:
\begin{align}
 &Q_{2a}(r) = P^{(2mL,\ |\ell| - \frac{1}{2})}_a(2r^2-1)\,,\\
 &Q_{2a + 1}(r) = r P^{(2mL,\ |\ell| + \frac{1}{2})}_a(2r^2-1)\,.
\end{align} 

Now we cast the differential operator $H_\ell$ to a rank-$n$ matrix $H_{ab}$, by projecting it to the Hilbert space spanned by the first $n$ elements of the basis \mbox{$\psi_a(r) = r^{|\ell|} (1-r^2)^{mL} \phi_a(r)$}. We have
\begin{equation}
\begin{split}\label{matrix elements}
 H_{ab} & = \bra{\psi_a}H_\ell\ket{\psi_b} = \int_0^1 \dd r\ r^{2|\ell|} (1-r^2)^{2 mL} \Bigg[
 \frac{1}{2}\(\phi_{a2} \phi'_{b1} - \phi_{a1} \phi'_{b2} - \phi'_{a2} \phi_{b1} + \phi'_{a1} \phi_{b2}\) + \\
 &+ \frac{\ell}{\alpha(r)}\(\phi_{a2} \phi_{b1} + \phi_{a1} \phi_{b2}\) + \frac{m}{\beta(r)} \(\phi_{a1} \phi_{b1} - \phi_{a2} \phi_{b2}\) + \Phi(r) \(\phi_{a1} \phi_{b1} + \phi_{a2} \phi_{b2}\)\Bigg]\,.
\end{split}
\end{equation} 

Even when using orthogonal polynomials, the basis $\psi_a$ turns out to be not orthogonal, because it involves linear combinations of polynomials of different degree. Therefore, it has a non-trivial overlap matrix
\begin{equation}
 G_{ab} = \braket{\psi_a}{\psi_b} = \int_0^1 \dd r\ r^{2|\lambda|} z^{2 m} \(\phi_{a1} \phi_{b1} + \phi_{a2} \phi_{b2}\)\,.
\end{equation} 

Given the matrices $H$, $G$, the approximate eigenfunctions of $H_\ell$ are obtained by solving the generalized eigenvalue problem
\begin{equation}
 H v=\omega G v\,.
\end{equation} 

Since $H$ is hermitian and $G$ is hermitian and positive definite, the eigenvalues $\omega$ are real and the vectors $v$ form a basis. The matrix $U$ that has the eigenvectors $v$ for columns satisfies\footnote{Note that $U$ is not unitary.}
\begin{align}
 & U^\star_{ia} G_{ab} U_{jb} = \delta_{ij}\,, && U^\star_{ia} H_{ab} U_{jb} = \omega_i \delta_{ij}\,.
\end{align} 
Therefore, $\omega_i$ are the approximate eigenvalues of the Dirac Hamiltonian $H_\ell$, and the corresponding approximate eigenfunctions are given by
\begin{equation}
 \psi_i(r) =  r^{|\lambda|} (1-r^2)^m U_{ia} \phi_a(r)\,.
\end{equation} 

The question remains of how to compute the matrix elements $G_{ab}$ and $H_{ab}$. By using the recursion relation for the polynomials $Q_a$,  it is possible compute analytically the following quantities
\begin{align}
 &h_a = (Q_a, Q_a) \,,&&K_{ab} = (Q_a, Q'_b) - (Q'_a, Q_b)\,,
\end{align} 
with which it is then possible to directly compute $G_{ab}$ and the matrix elements of the kinetic term of the Hamiltonian. The terms involving $\alpha$, $\beta$ and $\Phi$, instead can be reduced to the form $(Q_a, f(r) Q_b)$. We compute them \cite{position_operator} by expanding $Q_a$, $Q_b$ over approximate eigenfunctions of the position operator $r$, also called cardinal functions. These are polynomials $C_i$ such that
\begin{align}
 (C_i, C_j) = \delta_{ij}\,, && (C_i, r C_j) = r_i\delta_{ij}\,.
\end{align}
They can be obtained as linear combinations of the polynomials $Q_a$ by diagonalizing the matrix
\begin{align}
 R_{a b} = \frac{(Q_a, r Q_b)}{\sqrt{h_a h_b}}\,,
\end{align}
which can be computed analytically, again using the recursion relation. Let $V$ be the orthogonal matrix that diagonalizes $R$:  $V^\dag R V = r$. Then
\begin{equation}
 Q_a(r) = \sqrt{h_a} V_{ai} C_i(r)
\end{equation} 
and we have
\begin{equation}
 (Q_a, f Q_b) = \sqrt{h_a} U_{ai} (C_i, f(r) C_j) U_{bj} \sqrt{h_b}\,.
\end{equation} 
Now the operator $r$ inside $f$ is acting against an approximate eigenfunction, and we have
\begin{equation}
 (Q_a, f Q_b) \simeq \sqrt{h_a} U_{ai} f(r_i) U_{bi} \sqrt{h_b}\,.
\end{equation} 
This approximation becomes an equality if the function $f$ is a polynomial and the total degree of $Q_a Q_b f$ is less than $2n$.
Otherwise it is allows for an error, which is exponentially small in $n$, provided $f$ is analytic over the interval $[-1,1]$. Using this approximate integration, the matrix elements of $H_{ab}$ can be computed.

\subsection{Near-boundary singularity}
\label{near boundary section}

It is well known from the literature on Casimir energy \cite{unnatural_acts}, that quantum fields in spaces with a boundary have peculiar behavior. This issue is very relevant to the problem at hand, because AdS is a space with a boundary. It turns out that the boundary at $r = 1$ causes the currents to  approach their $s = 0$ profile in a non-uniform way, more or less like
\begin{equation}
 f(r, s) = r^{\frac{1}{s}}\quad \text{for $0 < r <1$}
\end{equation} 
approaches its limit $f(r, 0) = 0$.

\begin{figure}[ht]
\begin{center}
\includegraphics[width=\textwidth]{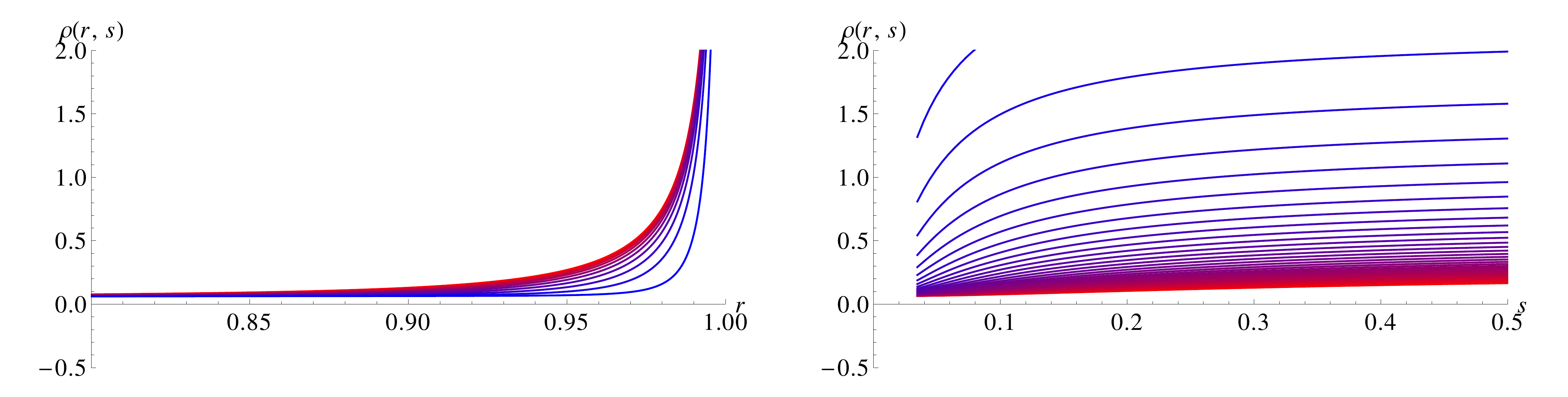}
\includegraphics[width=\textwidth]{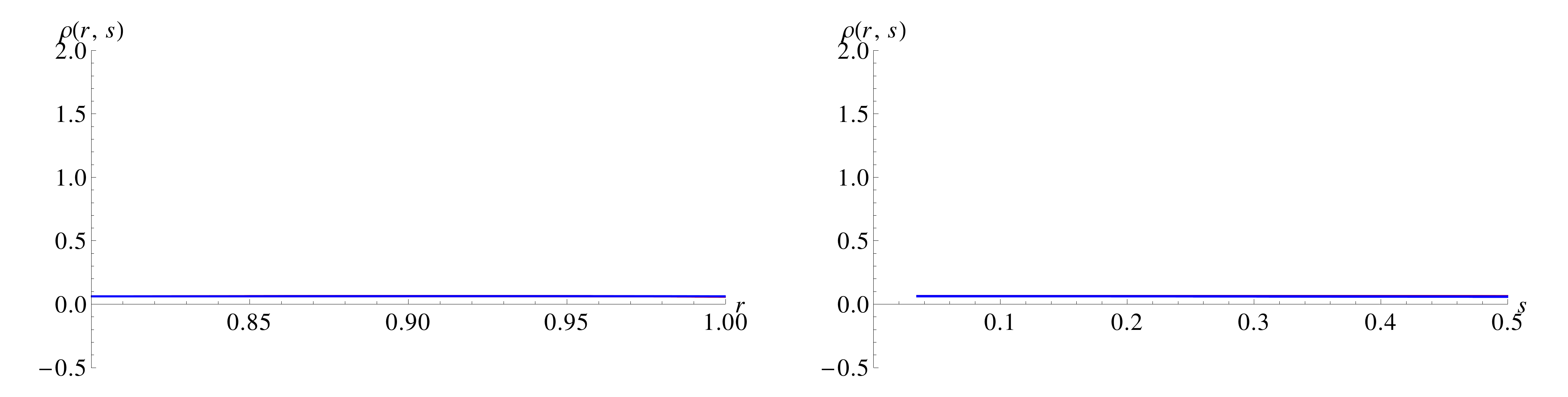}
\end{center}
\caption{\label{boundary divergence plot} \small Near-boundary behavior of the charge density, before (top) and after (bottom) the subtraction of a boundary counterterm.
On the left the radial ($r$) dependence is plotted, and the regulator ($s$) dependence
is encoded in the color; vice versa on the right.  After the subtraction 
the convergence in $s$ is completely uniform in $r$ and the extrapolation to $s=0$ 
is robust.}
\end{figure}

Figure \ref{boundary divergence plot}, top, exemplifies this phenomenon. It shows the charge density $\rho \equiv \beta^4 J^t$ in pure AdS geometry, with an electric potential\footnote{This potential has $\Phi'(1)\neq 0$, $\Phi''(1) = \Phi^{(3)}(1) = 0$.}
\begin{align}
 \Phi = \frac{V}{16}\(15 r^2 - 5r^4 + r^6\)\,.
\end{align}
The charge density is plotted at nonzero $s$, but after the adiabatic expansion has been subtracted, so it has a finite limit as $s \to 0$. On the left, $\rho$ is shown as a function of $r$, for several values of $s$, the bluer the smaller. On the right $\rho$ is shown as a function of $s$, for several values of $r$, the bluer the closer to the boundary. The $r$-profiles approach non-uniformly a limiting flat curve. We find that this kind of behavior is present whenever $\Phi'(1) \neq 0$. Intuitively, it can be explained as a layer of charge at $r = 1$ induced by the electric field at the boundary through vacuum polarization\footnote{This layer is a fictitious, finite-cutoff effect that disappears as $s \to 0$, but was mistaken for real in \cite{quantum_electron_star}.}.

Although the $s \to 0$ limit is well defined and finite, the fact that it is approached non-uniformly makes it extremely hard to reach in practice. In fact, the computational cost grows exponentially as $s$ decreases, because more and more wavefunctions must be retained to compute the bare currents. This would put beyond reach the computation of the currents near the boundary.

Fortunately, the contribution to the current that comes from the boundary conditions, and that vanishes non-uniformly at $s = 0$, can be computed analytically and subtracted. The plot at the bottom of fig.~\ref{boundary divergence plot} shows $\rho$ after the subtraction of
\begin{equation}\label{boundary counterterm}
 \Delta \rho = \frac{\Phi'(1)}{6 \pi ^2 s}\[ 3\left(1+4 \frac{z^2}{s^2} \right) \tan ^{-1}\left(\frac{s}{2 z}\right)- \frac{s}{z} \left(1+6 \frac{z^2}{s^2}\right)\]\,,\quad z = 1 - r\,:
\end{equation} 
the non-uniform singularity has been removed. This counterterm correctly accounts for the finite-$s$ effect of the boundary conditions, without altering the $s \to 0$ limit, because it vanishes at $s = 0$, for any $r\in[0,1)$. After this subtraction, the limit $s\to0$ is easily and safely taken by extrapolation.

There are also non-uniform singularities in the stress tensor, when $\alpha^{(3)}(1)\neq 0$ or, for $m\neq 0$, singularities proportional to $\beta'(1)$ and $\beta^{(3)}(1)$. For each of these singularities, a boundary counterterm like (\ref{boundary counterterm}) must be and has been derived.

These counterterms can be obtained using an approach similar to the adiabatic expansion, assuming that the length scale over which the background varies is much larger than the both the distance $z = 1 -r$ from the boundary and the point splitting separation $s$. Using the simple example
\begin{equation}
 \[-\partial_z^2 -\nabla^2 + V(z, x)\] G(z, x; z', x') = \delta(z - z')\delta(x - x')\,,
\end{equation}
we can expand for small $z$ and small $x - x'$, and write the previous equation as
\begin{equation}
 \[G_0^{-1} + A\] G = \onematrix\,,
\end{equation} 
with
\begin{align}
 &G_0^{-1} = -\partial_z^2 -\nabla^2 \,, \\
 &A = V(0, x') + V_{,z}(0, x') z + V_{,i}(0, x') (x - x')_i + \ldots\,.
\end{align}  

The Green's function is given by
\begin{equation}\label{G series}
 G = G_0 \sum_{n = 0}^{\infty} (- A G_0)^n\,,
\end{equation} 
but, in this case, $G_0$ must account for the boundary conditions on the fields at $z = 0$. For example, for Dirichlet boundary conditions:
\begin{equation}
 G_0(z, x; z', x') = 2 \int \dbar q \dbar^{d-1} p\ \frac{\sin(q z) \sin(q z') e^{i p(x - x')}}{q^2 + p^2}\,.
\end{equation} 

Because there is no translational invariance, it is better to stay in position space in the $z$ direction, so we write
\begin{equation}
 G_0(z, x; z', x') = \int\dbar^{d-1} p\ G_p(z, z') e^{i p(x - x')}\,,
\end{equation} 
where
\begin{equation}
 G_p(z, z') =
\begin{cases}
 -\frac{1}{p}\sinh(pz) e^{-p z'}&\text{for $z < z'$}\\
 -\frac{1}{p}\sinh(pz') e^{-p z}&\text{for $z > z'$}\,.
\end{cases}
\end{equation} 

Let us consider the contribution of the term $V_{,z}(0,x')$ to the Green's function $G$. We have
\begin{equation}
 \Delta G(z,x;z',x') = -\int\dbar^{d-1}p\ e^{ip(x-x')} \int_0^{\infty} \dd \zeta \ G_p(z,\zeta) V_{,z} \zeta G_p(\zeta,z')\\
\end{equation} 

For simplicity, we let $z' = z$ and we have
\begin{equation}
\begin{split}
 \Delta G(z,x;z,x') & = -V_{,z}\int\dbar^{d-1}p\ e^{ip(x-x')} \frac{z}{4p^3} \[1- \(1 + p z\)e^{-2pz}\]\\
& = \frac{V_{,z} z}{4 \pi ^2} \left[\log \left(\frac{4 z^2}{s^2}+1\right)+2 \frac{z}{s} \tan
   ^{-1}\frac{s}{2 z}\right]\,,
\end{split}
\end{equation} 
where the last result is specific to $d = 4$, and where $s = |x - x'|$.

If we expand this expression at small $s$, we find the divergent term
\begin{equation}
 \Delta G(z,x;z,x')\sim -\frac{V_{,z}z}{4 \pi ^2} \log\frac{s^2}{4z^2}\,.
\end{equation} 
This term was already obtained from the adiabatic expansion. It is the second term in (\ref{sample adiabatic expansion}), with $m^2$ replaced by $V_{,z}z$. In fact, $V(z)$ is locally a mass term $m^2(z)$, and the current expansion is reproducing the divergence $m^2(z) \log s$ through its series expansion about $z = 1$.

If we subtract both this logarithmic divergence and the $\mathcal O(s^0)$ term, we obtain a quantity that vanishes as $s \to 0$:
\begin{equation}
 \Delta G(z,x;z,x') = \frac{V_{,z} z}{4 \pi ^2}  \left[\log \left(1+\frac{s^2}{4 z^2}\right)+ 2 \frac{z}{s} \tan^{-1}\frac{s}{2 z}-1\right]\,,
\end{equation} 
and which would be a boundary counterterm for the coincidence limit of the Green's function. Unfortunately, this example is limited, in that this expression vanishes uniformly, so this counterterm is not necessary. However, by the same means but more complicated algebra, it is possible to compute the counterterm (\ref{boundary counterterm}), which instead is of  crucial importance.

\section{Solution of Einstein's equations}
\label{sec:einstein}

Given an electromagnetic current and a stress tensor with 
enough symmetries (`co-homogeneity one'), the solution of Einstein and Maxwell equations 
is a relatively standard problem. Let us briefly describe the method we used.

With the ansatz
\begin{align}
\label{eq:metricansatz}
 & g = \frac{1}{\beta^2(r)}\( -\dd t^2 + \dd r^2 + \alpha^2(r) \dd \Omega_2^2 \)\,,  && A = \Phi(r) \dd t\,,
\end{align}
the equations
\begin{equation}
\left\{
\begin{aligned}
  &D_\mu F^{\mu \nu} = q^2 J^\nu\\
  &G_{\mu\nu} + \Lambda g_{\mu \nu} = \kappa^2 T_{\mu\nu}\,,
\end{aligned}\right.
\end{equation} 
become
\begin{equation}
\left\{
\begin{aligned}
  &\beta^4\(\Phi'' + \frac{2\alpha'}{\alpha} \Phi'\) &&= q^2 J^t\\
  &\beta^2\(\frac{2 \alpha ''}{\alpha }-\frac{4 \alpha' \beta '}{\alpha \beta }+\frac{\alpha'^2-1}{\alpha^2}-2 \frac{\beta''}{\beta} + 3 \frac{\beta'^2}{\beta^2}\)& - \frac{3}{L^2} & = \kappa^2 T^{t}{}_t\\
&\beta^2\(-\frac{4 \alpha ' \beta '}{\alpha \beta}+\frac{\alpha'^2 - 1}{\alpha ^2}+3 \frac{\beta '^2}{\beta^2} \) &- \frac{3}{L^2} & = \kappa^2 T^{r}{}_r\\
&\beta^2\(\frac{\alpha ''}{\alpha }-\frac{2 \alpha' \beta'}{\alpha \beta} - 2\frac{\beta''}{\beta} + 3 \frac{\beta'^2}{\beta^2}\)&- \frac{3}{L^2}& = \kappa^2 T^{s}{}_s\,,
\end{aligned}\right.
\end{equation} 
where $L$ is the radius of the asymptotic AdS geometry, that is $\Lambda = -3 / L^2$. Because of spherical symmetry, we have  $T^\theta{}_\theta = T^\phi{}_\phi$, and we defined $T^{s}{}_s \equiv T^\theta{}_\theta = T^\phi{}_\phi$.

The first equation is simply Gauss' law; it is a linear equation, and does not require further discussion. The three Einstein's equations are not independent, because the Einstein tensor is covariantly conserved, that is, $D_\mu G^\mu{}_\nu = 0$ identically. This fact constrains the stress tensor to be convariantly conserved too:
\begin{equation}
 T^r{}_{r,r} + \(2\frac{\alpha'}{\alpha}-3\frac{\beta'}{\beta}\) T^r{}_r + \frac{\beta'}{\beta} T^t{}_t + 2\(\frac{\beta'}{\beta} - \frac{\alpha'}{\alpha}\)T^s{}_s = 0\,,
\end{equation} 
and this reduces the independent components from three to two.

We demand that $\alpha(0) = 0$ (regularity in IR) and $\beta(1) = 0$ (asymptotically AdS in UV). This sets the coordinate location of the center of the space ($r = 0$) and the boundary ($r = 1$). It is useful to expand the equations near these two points, to understand the asymptotic behavior of the sections. In order to do so, some knowledge of the behavior of the stress tensor is needed, which can be inferred by computing them explicitly in few sample backgrounds. Based on this, we can assume that the stress tensor is analytic at the boundary, and that $T^{t}{}_{t}(1) = T^{r}{}_{r}(1) = T^{s}{}_{s}(1)$. This is due to the symmetry of AdS space, which forces the stress tensor at the boundary to be simply a correction to the cosmological constant. This correction comes from the high energy modes, and hence is independent of $\Phi$, so we absorb it directly into $\Lambda$. The next three derivatives vanish, so we have
\begin{align}
\label{eq:nearboundarystress}
 & \kappa^2 T^t{}_t =  t_4\, (1-r)^4 + \mathcal{O}(1-r)^5\\
 & \kappa^2 T^s{}_s =  s_4\, (1-r)^4 + \mathcal{O}(1-r)^5
\end{align}
and
\begin{align}
\begin{split}
 &\alpha(r) = a_0 -\frac{(1-r)^2}{2a_0}  + a_3 (1-r)^3+\\
 &\phantom{\alpha(r) \sim} +\[\frac{a_0 L^2}{4} (t_4 - s_4) + \frac{1}{24a_0^3}\](1-r)^4 + \mathcal{O}(1-r)^5\,,
\end{split}\\
\begin{split}
 &\beta(r) = \frac{1-r}{L} - \frac{(1-r)^3}{6 a_0^2 L} + b_4 (1-r)^4+\\
 &\phantom{\beta(r) \sim} +\[\frac{L}{10}(t_4 - 2s_4) + \frac{1}{120a_0^4L}\](1-r)^5 + \mathcal{O}(1-r)^6\,.
\end{split}
\end{align} 

At the center we have
\begin{align}
 &\alpha(r) = r - \frac{1}{6b_0^2} \(\frac{1}{L^2} + \kappa^2T^s{}_s(0) -\frac{2}{3} \kappa^2T^t{}_t(0) \) r^3 + \mathcal O (r^5)\,,\\
 &\beta(r) = b_0 - \frac{1}{2b_0} \(\frac{1}{L^2} + \frac{1}{2} \kappa^2 T^s{}_s(0) -\frac{1}{6} \kappa^2 T^t{}_t(0) \) + \mathcal O (r^4)\,.
\end{align} 
Moreover, since the equations are symmetric under $r \to -r$, we can take $\alpha$ to be odd and $\beta$ and $\Phi$ to be even, provided that the currents are also even. Since the currents are even when the sections have definite parity (see (\ref{parity})), this assumption is self-consistent. 

The constants $a_0$, $a_3$, $b_0$ and $b_4$ are not fixed by the series expansion. They are four integration constants, which take a precise value in the unique solution that matches the two expansions at the edges. The constant $a_0$ is related to the radius of the sphere of the boundary theory, whereas $a_3$ and $b_4$ are related to the expectation value of the boundary stress tensor \cite{Balasubramanian:1999re, Skenderis:2002wp}, as described in section \ref{sec:results} (see \eqref{eq:bdystress}).

On a more practical level, we solve the equations using spectral methods. We represent the sections $\alpha$, $\beta$ and $\Phi$ as polynomials of moderate degree. There is some freedom in choosing what basis to use for the space of polynomials. We use Chebyshev polynomials of the appropriate parity as starting point, and we found important to take linear combinations, so that each basis element satisfies
\begin{align}
 \alpha'(1) = 0\,, && \beta(1) = 0\,, &&\beta''(1) = 0\,, &&\Phi(1) = 0\,.
\end{align}
The condition $\beta''(1) = 0$ is particularly important for the stability of Newton's method.

\section{Results and Discussion}
\label{sec:results}

Let us briefly recapitulate the setup. We are considering a quantum fermionic field in interaction with classical gravity and a classical U$(1)$ gauge field, in asymptotically anti-de Sitter spacetime. The class of backgrounds we are considering is described by the following ansatz\footnote{These coordinates are related to the coordinates in (\ref{the background}) by a rigid rescaling $t \to z_m t$, $z \to z_m(1 - r)$. The discussion of the results is slightly more transparent in these coordinates.}
\begin{align}
  g = \frac{1}{\beta^2(z)}\left[- \dd t^2 + \dd z^2  + \alpha^2(z)\dd \Omega_2^2\right]\,,&& A = \Phi(z) \dd t\,.
\end{align} 
The AdS boundary is at $z = 0$, where the conformal factor $\beta(z) \sim z / L$, and $L$ is the radius of the asymptotic AdS geometry. We assume that spacetime ends smoothly at $z = z_m$, and the spatial sections can be visualized as 3-balls with center at $z = z_m$ and edge at $z = 0$. Global AdS is a metric of this class, with $\beta = \sin {z\over L}$, $\alpha = L \cos {z \over  L}$, $z_m = \pi{L \over  2}$.

From the dual point of view, we are considering a 2 + 1 dimensional conformal field theory, defined on a sphere of radius $R = \alpha(0)$. This CFT has a global U$(1)$ symmetry, for which we turn on a chemical potential $\mu = \Phi(0)$, and a fermionic operator charged under the U$(1)$ symmetry, whose correlation functions we wish to study.

The model depends of four dimensionless parameters $q$, $\kappa/L$, $m L$, $\mu R$. The U$(1)$ coupling $q$, the gravity coupling $\kappa/L$ and the fermion mass $m L$ should be thought as parametric labels 
(like the number of species of fields)
specifying the dual CFT. Then, for a given CFT, dimensionless quantities depend on $\mu$ and $R$ only through the combination $\mu R$.

The duality allows the computation of several CFT quantities, particularly the U$(1)$ charge density $\rho_b$, the energy density $\epsilon_b$, the pressure $p_b$, and the fermion spectral function. The thermodynamic responses are given by\footnote{The result for $\epsilon_b$ and $p_b$ requires holographic renormalization as described in \cite{Skenderis:2002wp}.}
\begin{align}
 &\rho_b = - \Phi'(0)\,,\\
 \label{eq:bdystress}
 &\epsilon_b = \frac{1}{R}\alpha^{(3)}(0) - \frac{L}{3}\beta^{(4)}(0)\,,\\
 &p_b = -\frac{1}{2R}\alpha^{(3)}(0) + \frac{L}{3}\beta^{(4)}(0)\,.
\end{align}

For what concerns the spectral function, it is important to notice that the CFT is defined on a sphere, and hence 
the spectrum of the many-body Hamiltonian is discrete, and single particle states are labeled by the partial wave number $\ell$. Consequently, the fermion spectral function
\begin{equation}
 A( \ell, \omega) = \sum_\alpha \[ \left|\bra{\alpha} c^\dag_{ \ell} \ket{\text{gd}}\right|^2 \delta(\omega - E_\alpha) + \left|\bra{\alpha} c_{ \ell} \ket{\text{gd}}\right|^2 \delta(\omega + E_\alpha)\]
\end{equation}
is composed of a discrete set of delta functions that track the many-body eigenvalues. However, for $\mu R \gg 1$ the effect of the infrared regulator $R$ becomes negligible, and the flat space spectral function is recovered. In fact, in this regime, one can identify $\ell / R$ with a continuous momentum label $k$, and the delta functions merge into a continuum. Hence, the regime $\mu R \gg 1$ is of the greatest interest. Holographically, the location of the delta-functions in the $k$-$\omega$ plane coincides with the spectrum of the bulk Dirac Hamiltonian.

\begin{figure}[t!]
\begin{center}
\includegraphics[width=0.49\textwidth]{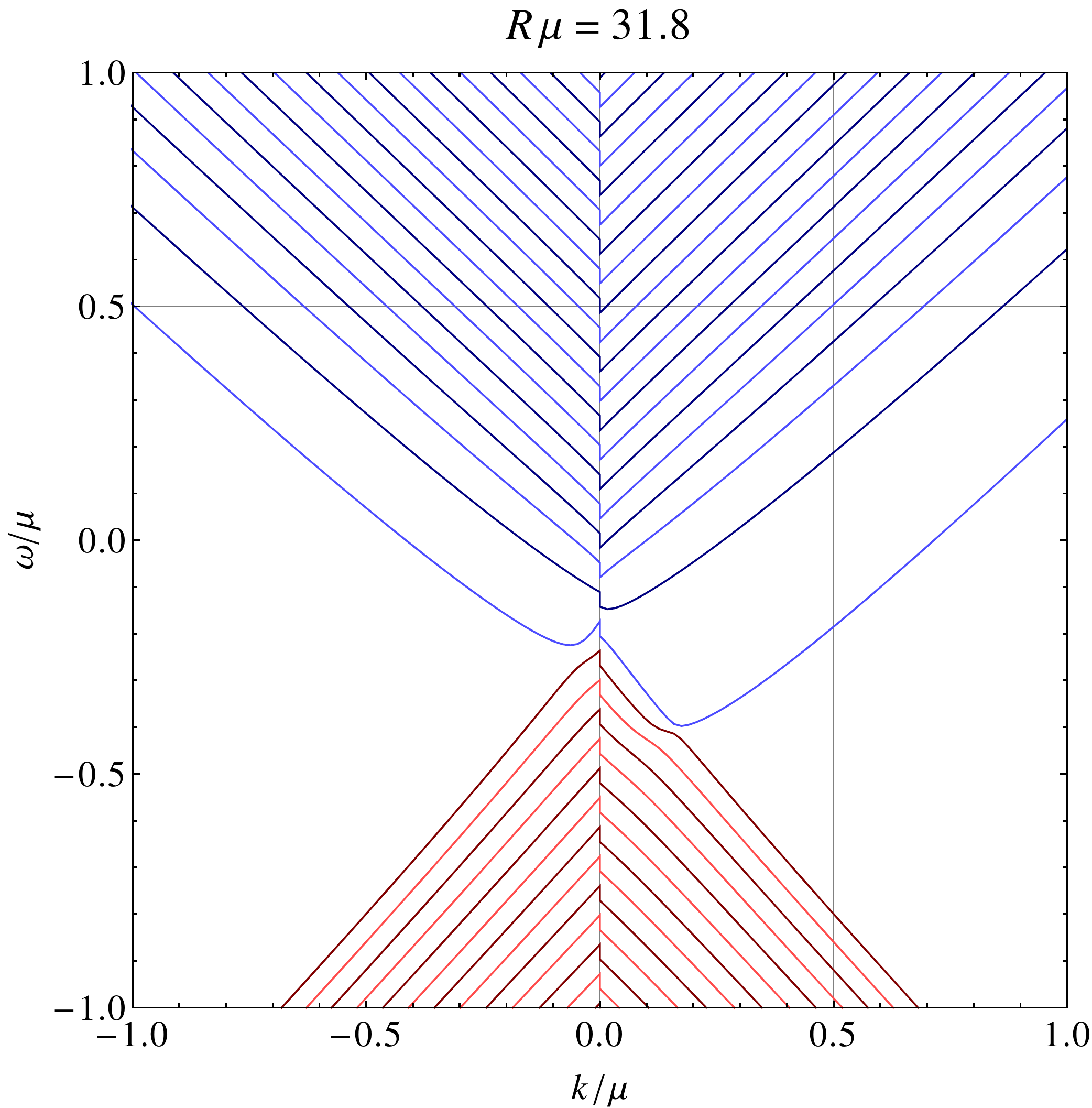}
\includegraphics[width=0.49\textwidth]{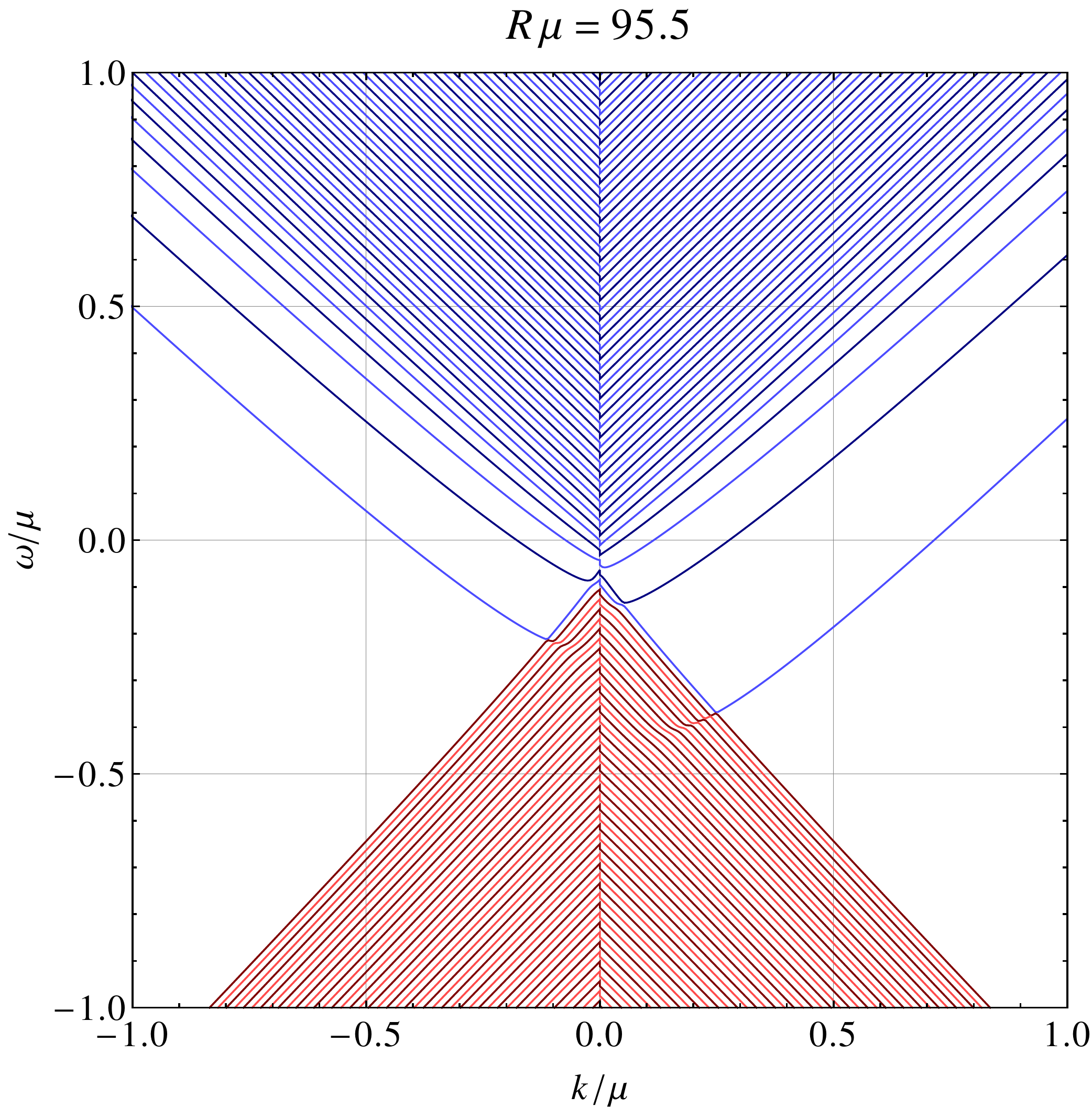}
\end{center}
\caption{\small \label{spectral function no backreaction}Example of spectral function: $\kappa^2 / L^2 = 0.0$, $q^2 = 4.0$, $mL = 0.0$. Compare with figure 3 of \cite{quantum_electron_star}. Only discrete values of $k = \ell / R$ carry spectral weight, but here lines are shown as if $k$ were a continuous variable. The difference is negligible as $\mu R \to \infty$.}
\end{figure}

As an example of how the continuum is approached, consider figure \ref{spectral function no backreaction}. The plots refer to a frozen global AdS geometry, with self-consistently determined gauge field, and display the location of the delta function peaks of the spectral function. 
As $\mu R$ increases, a continuum emerges in the light cone $\omega^2 > k^2$, outside of which a number of isolated bands remain. These results agree with our previous findings in \cite{quantum_electron_star}, but here they have been derived with far greater care for all the regularization and renormalization issues, therefore giving an important check of the correctness of our previous work. For an interpretation see \cite{quantum_electron_star}.

\begin{figure}[t!]
\begin{center}
\includegraphics[width=0.8\textwidth]{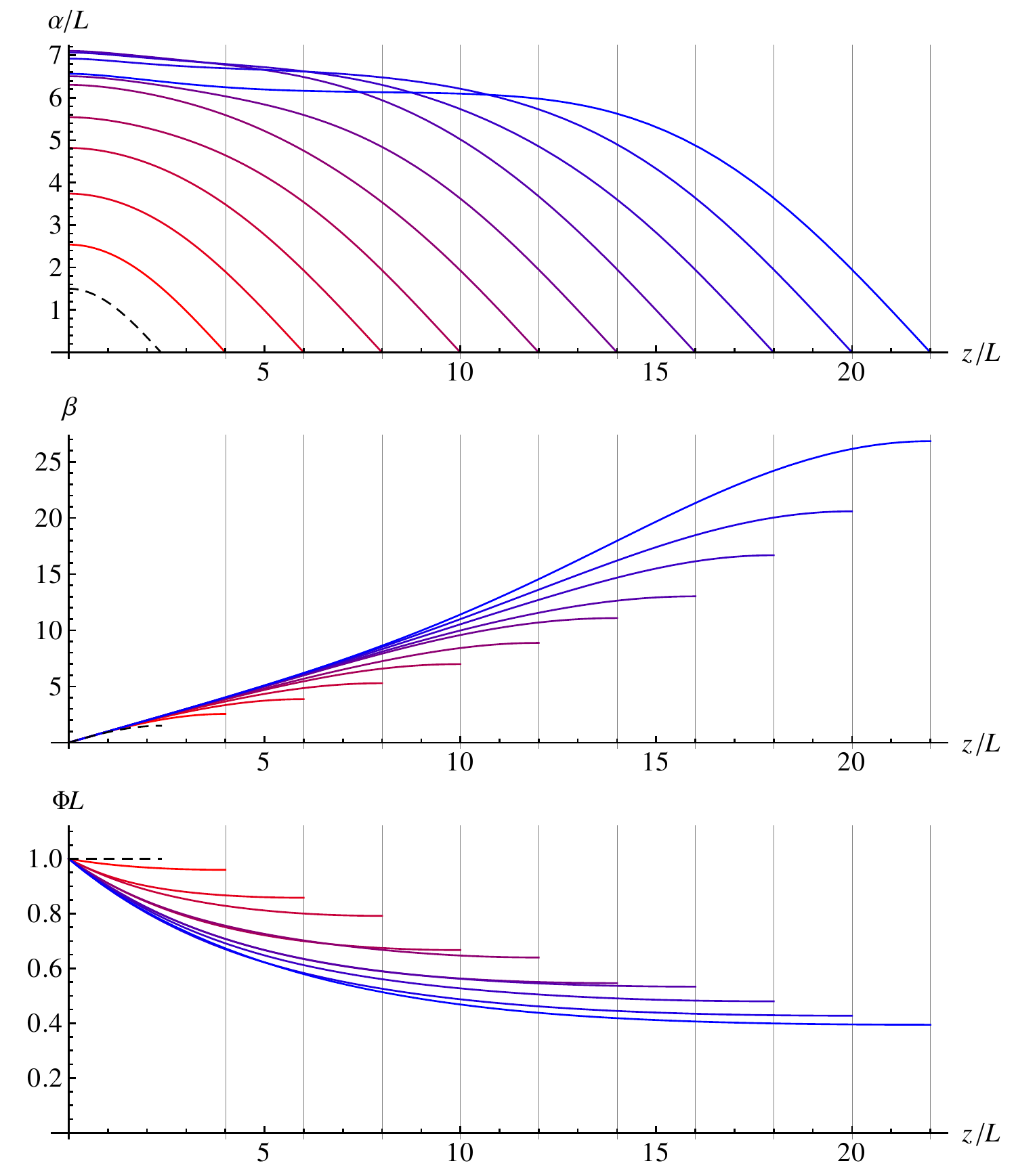}
\end{center}
\caption{\small \label{backgrounds plot} Self consistent profiles for $\kappa ^2/L^2 = 0.1$, $q^2 = 1.0$, $m L = 0.0$}
\end{figure}

Moving to the more interesting case of a dynamic metric, figure \ref{backgrounds plot} shows the profiles for $\alpha(z)$, $\beta(z)$ and $\Phi(z)$ in a set of self-consistent solutions. Without loss of generality, we set $\mu L = 1$. Then, by dialing $z_m$, we are able to vary the radius $R$ of the boundary sphere. If  $R < {3\over 2} L$, the fermions do not contribute any charge or energy density, and hence the background is given by global AdS geometry, with constant electric potential
\begin{align}
  g = \frac{L^2}{R^2\sin^2\frac{z}{R}}\left[- \dd t^2 + \dd z^2  + R^2 \cos^2 \frac{z}{R}\dd \Omega_2^2\right]\,,&& A = \frac{1}{L} \dd t\,.
\end{align} 
This is because the spectrum of the Dirac Hamiltonian in global AdS is discrete and gapped, the lowest positive-energy state being $\omega_0 = {3\over 2R}$. If the electric potential is smaller than this threshold, no charge is induced in the bulk. From the dual point of view, the infrared regulator opens a gap of order $1/R$ in the spectrum of charged excitation, and hence the system is incompressible for sufficiently small $R$. The critical solution $R = 3/2$ is shown with a dashed line in figure \ref{backgrounds plot}.

As $R$ goes beyond the critical value, the fermions start contributing nonzero charge and energy density, which then backreacts on the gauge field and the geometry. The expectation is that, at large $R$, the background would approach some asymptotic, $R$ independent behavior, at least for $z \ll z_m$. What we discover instead, is that rather soon $R$ stops growing as $z_m$ is increased and, more or less at the same point, the iterative algorithm becomes unstable and fails to converge. A notable feature of this instability is that, iteration after iteration, the value $\beta(z_m)$ grows beyond bounds, suggesting that the system would like to develop a horizon in the interior.

\begin{figure}[t!]
\begin{center}
\includegraphics[width=0.49\textwidth]{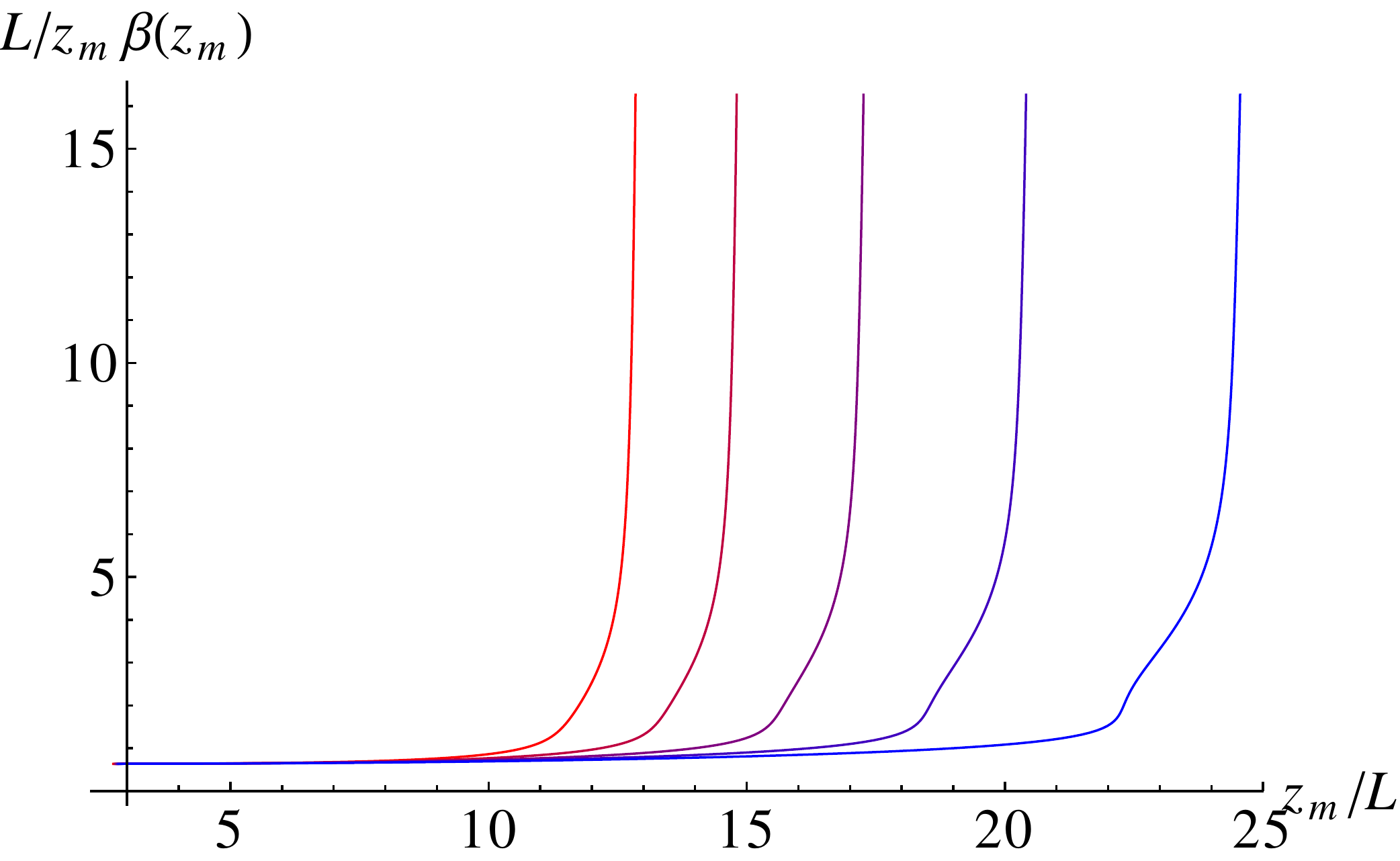}
\includegraphics[width=0.49\textwidth]{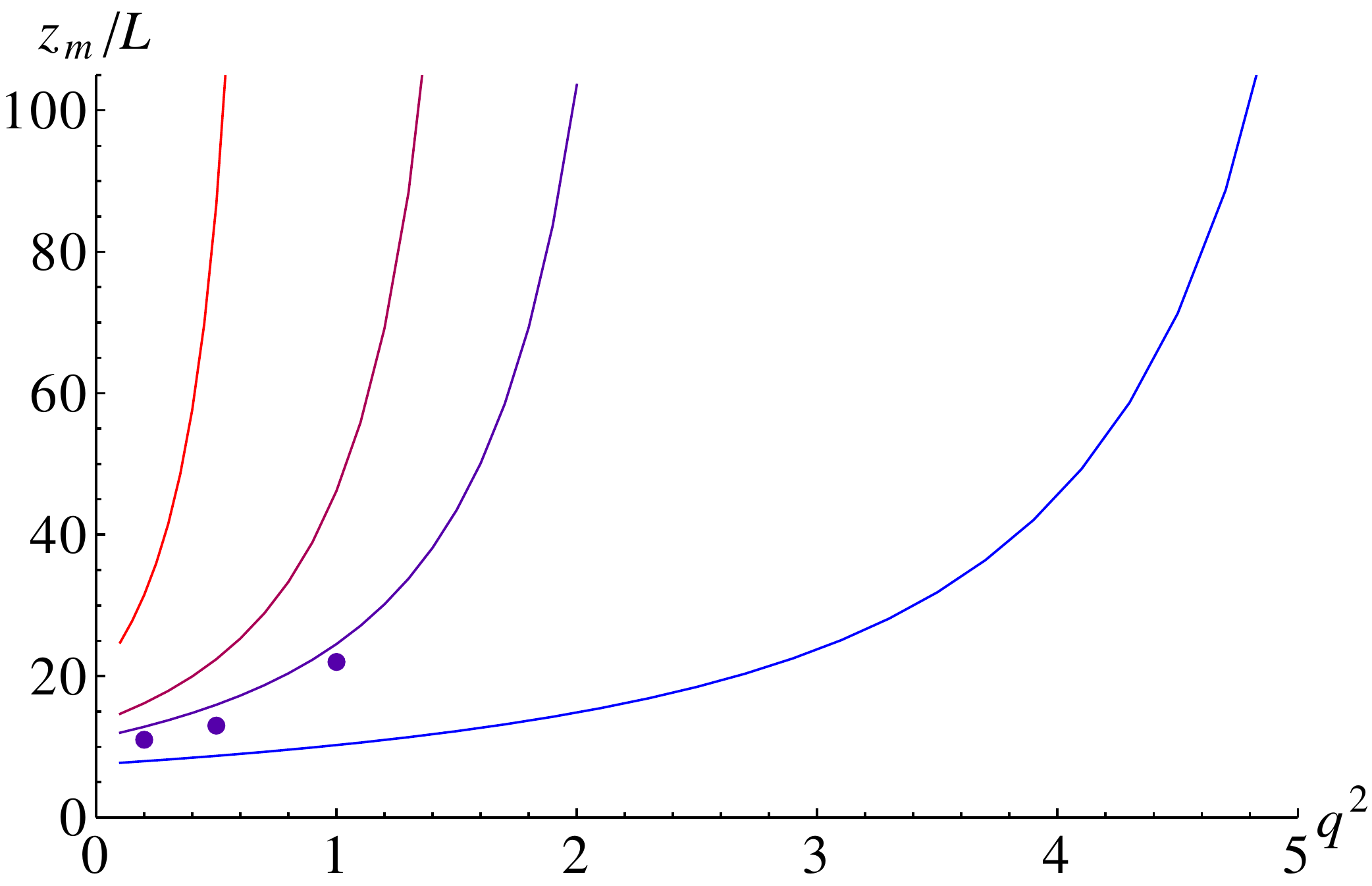}
\end{center}
\caption{\small \label{TF instablity} Location of the phase transition in the Thomas-Fermi approximation. On the left $\beta(z_m)$ against $z_m$. From red to blue $q^2 = 0.2,\,0.4,\,0.6,\,0.8$,  $\kappa^2/L^2 = 0.1$, , $m L = 0.0$. On the right, the critical $z_m$ as a function of $q^2$. The points show the approximate location of the numerical instability. From red to blue, $\kappa^2/L^2 = 0.01,\,0.05,\,0.1,\,0.5$, $m L = 0.0$.}
\end{figure}

Close to the instability, the present method of solution is not reliable enough  to determine whether there is an actual singularity or merely an algorithmic problem, but it gives some indications that the first option is the correct one. Therefore, to investigate better the issue, we solved the problem in the same setup, but within the Thomas-Fermi approximation. The results are shown in figure \ref{TF instablity}. It is manifest that $\beta(z_m)$ diverges at a finite value of $z_m$, over all the range of $q^2$ and $\kappa^2 / L^2$ that we have explored, indicating the presence of an actual singularity. The points on the right plot display the approximate location of the onset of the instability of the iterative algorithm, and agree quite well with the phase boundary determined by the Thomas-Fermi approximation.
 
Based on these results, we conclude that our model develops physical singularity as $z_m$ increases, such that $\beta(z_m)$ diverges at a finite value of $z_m$. While this phase transition may be interesting in itself, it precludes the possibility of studying the large $\mu R$ limit with the current method of solution. In fact, the geometry at $\mu R \gg 1$ is likely to have a zero temperature horizon or some other singularity in the interior, and this would cause the Dirac operator to be non-compact. While it may be possible to generalize the current method of solution to a Hamiltonian with a continuous spectrum, it is beyond the scope of the present article.

\begin{figure}[t!]
\begin{center}
\includegraphics[width=0.32\textwidth]{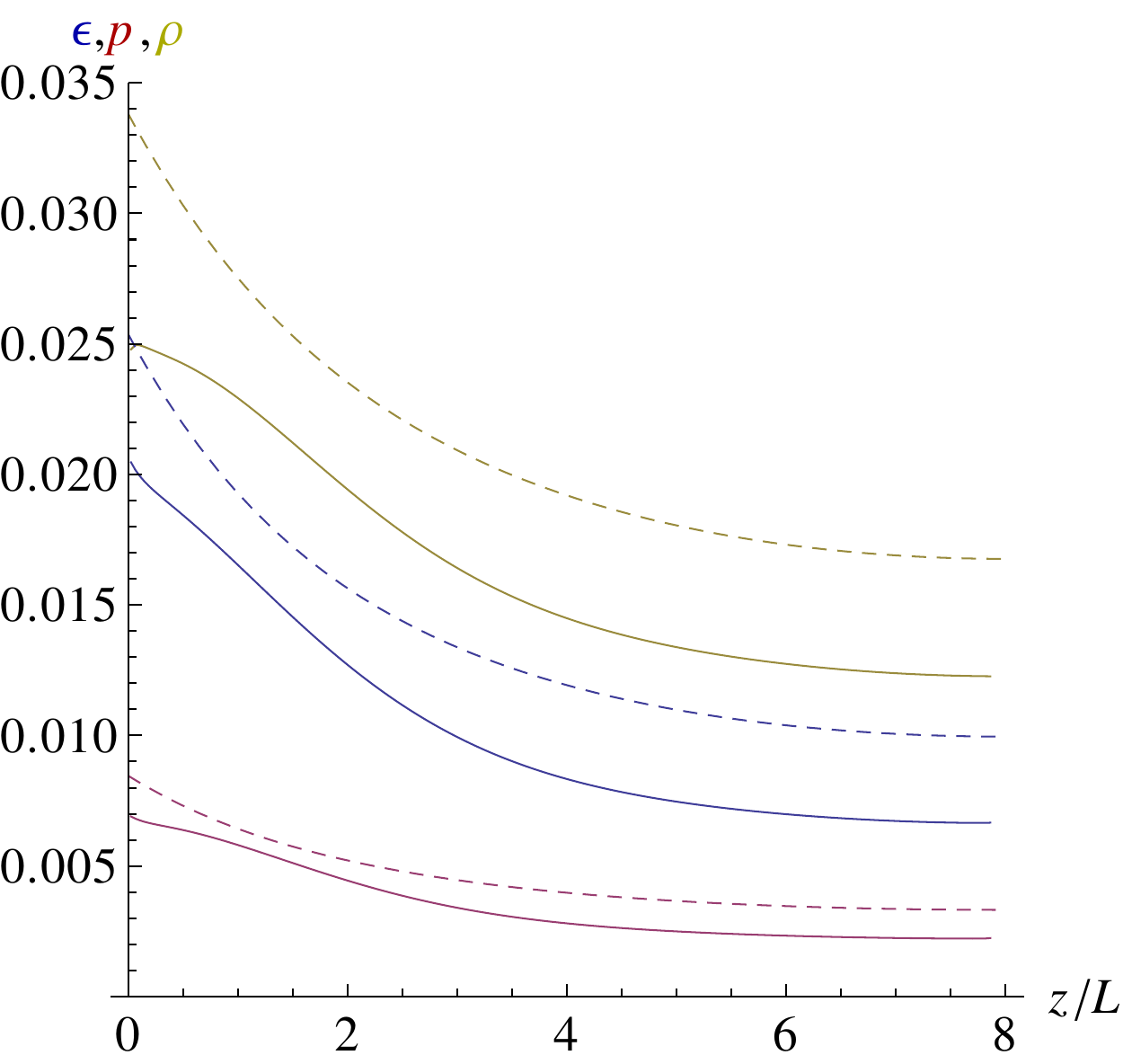}
\includegraphics[width=0.32\textwidth]{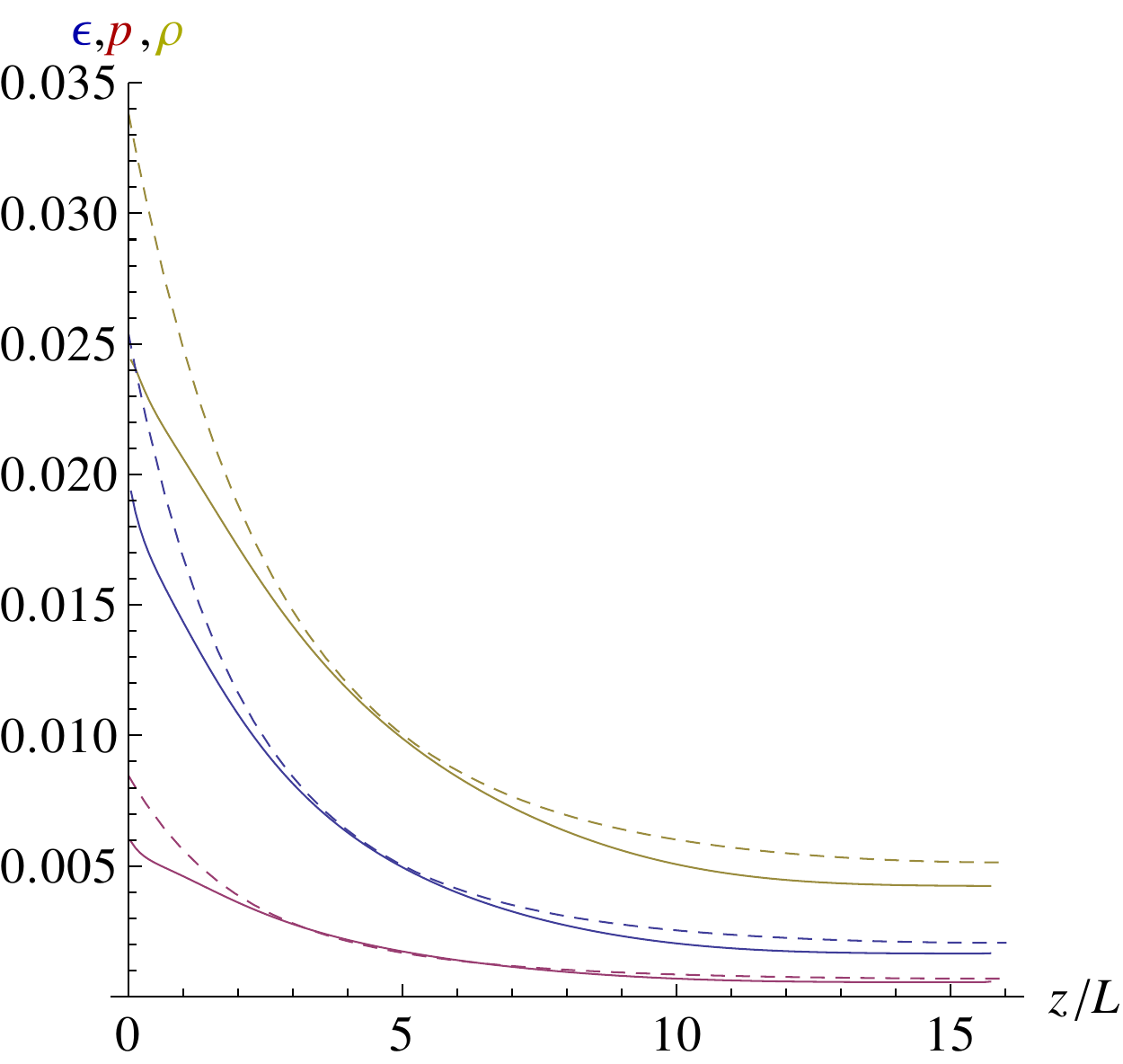}
\includegraphics[width=0.32\textwidth]{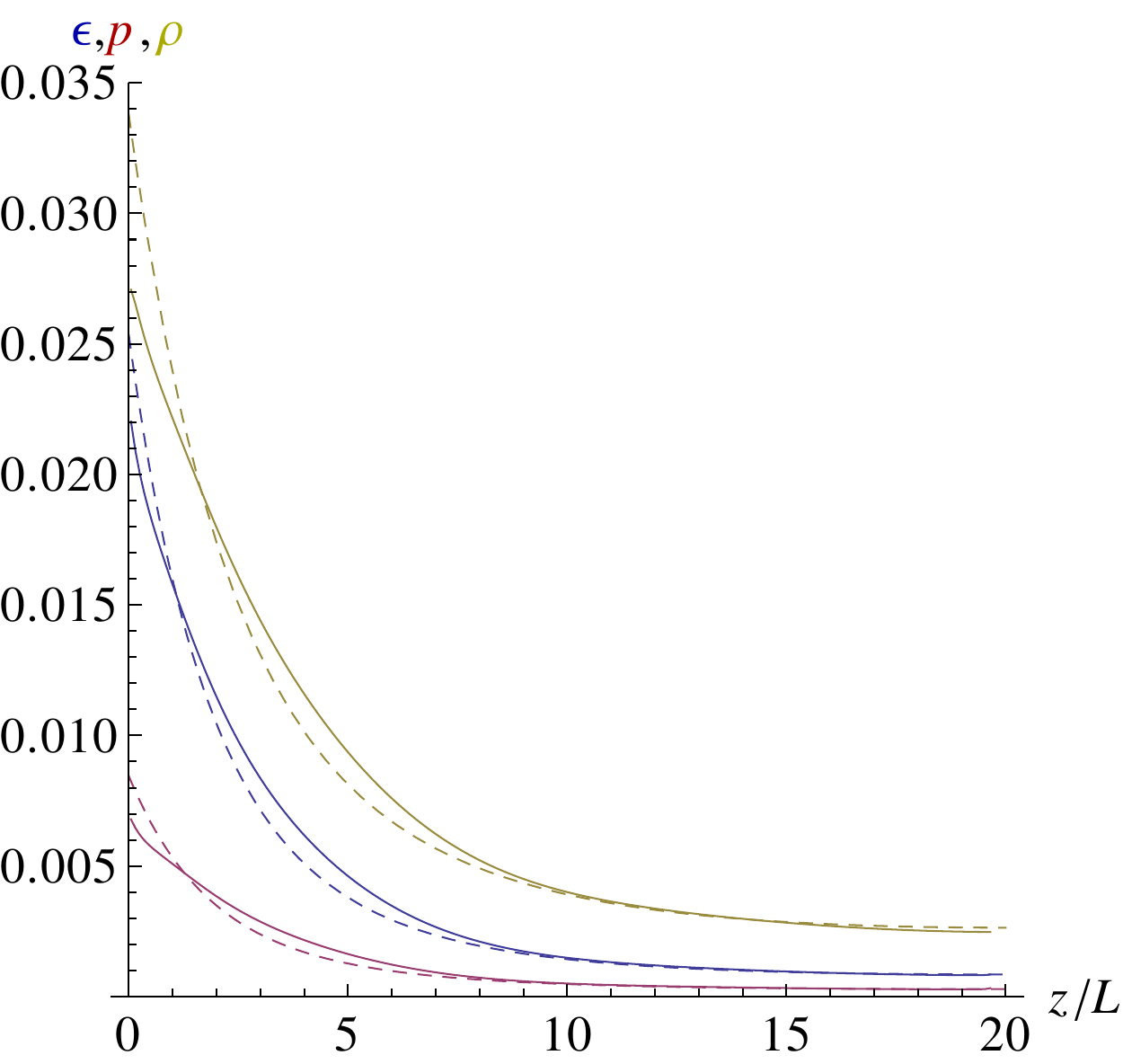}
\end{center}
\caption{\small \label{TF comparison} Comparison of the Thomas-Fermi approximation (dashed line) against the exact answer (solid line). $q^2 = 1.0,$,  $\kappa^2/L^2 = 0.1$, $m L = 0.0$. From left to right $z_m / L = 8,\, 16,\, 20$.}
\end{figure}

The accuracy with which the Thomas-Fermi approximation is able to predict the location of the numerical instability raises questions on its regime of validity. According to \cite{electron_star}, the Thomas-Fermi approximation is justified in the regime $m L \gg 1$, but we find it to be a good description even at $m L = 0$. In figure \ref{TF comparison} we compare the currents computed within the Thomas-Fermi approximation against the exact ones, in a self consistent background. It is apparent that, as $z_m$ grows, the Thomas-Fermi approximation becomes better. On the other hand, the approximation should be consistent if locally 
\begin{equation}
 \frac{|\nabla k_F(x)|}{k_F^2(x)} \ll 1\,,
\end{equation} 
which in the current setup ($m L = 0$) translates to
\begin{equation}
\frac{1}{\beta \Phi^2} \frac{d}{dz} (\beta \Phi) \ll 1\,.
\end{equation} 
This condition breaks down if $\beta$ diverges, as we have seen happens at a finite critical value of $z_m$. Near this critical value, it is not possible to reliably compute the exact answer, so we have no way of verifying this breakdown. We may summarize our present understanding of the Thomas-Fermi approximation by saying that it is inadequate at small $z_m$, it improves at larger values, and probably breaks down near the critical $z_m$.

We end this paper with a discussion of possible future directions.
\begin{itemize}

\item The phase transition that we observe in the Thomas-Fermi approximation certainly deserves further investigation, and currently lacks a clear holographic interpretation. We emphasize that it is a zero temperature phase transition different than the one described in \cite{Hartnoll:2010xj}.
It appears to be a confinement-deconfinement transition 
\cite{Witten:1998qj, Witten:1998zw}
driven by the fermion density.
Perhaps-related transitions between bulk solutions where all the charge is manifest in the geometry
and `fractionalized' solutions where electric flux emerges from the IR boundary 
were studied in \cite{Hartnoll:2011pp}. 
Those transitions differ from the one found here in that 
they were triggered by varying a scalar coupling,
and in that they occurred in the infinite-volume limit. 

\item It would be very important to develop a better understanding of the regime of validity of the Thomas-Fermi approximation.

\item In order to explore the large $\mu R$ regime, which is of the greatest physical interest, it may be necessary to generalize the methods presented in this paper to situations in which the Hamiltonian is not a compact operator. On the other hand, it may be sufficient to replace the radius of the sphere with some other infrared regulator that is better behaved.

\item To gain complete control over the approximations made in solving the model, it would be important to estimate the amplitude of the fluctuations of the metric and the gauge field, which have been neglected (in the Hartree-Fock approximation). This amounts to computing the current-current correlators. The same information could also be used to substitute Newton's method to the current naive iteration algorithm, allowing for a faster and more stable solver. This would probably allow to explore values of $z_m$ closer to the critical point.

\item In holographic duality, we are accustomed to interpreting the 
radial dependence of (bosonic) bulk fields 
as encoding running couplings 
in the dual QFT (along with some information
about the quantum state).
The holographic interpretation of the quantum
state of the bulk fermion fields
poses an interesting question of principle
for holographic duality.
It appears to provide a concrete example
of the `quantum renormalization group'
described in \cite{Lee:2010ub, Lee:2013dln, Lee:2012xba}.

\end{itemize}

\vfill\eject
\vskip.2in
{\bf Acknowledgements}
We thank 
Tom Banks,
Simon Gentle,
Sean Hartnoll, 
Julius Kuti,
Simon Ross, 
Subir Sachdev,
and Brian~Swingle,
for discussions, comments and encouragement.
This work was supported in part by
funds provided by the U.S. Department of Energy
(D.O.E.) under cooperative research agreement DE-FG0205ER41360,
in part by the Alfred P. Sloan Foundation,
and in part by DOE-FG03-97ER40546.
Simulations were done on the MIT LNS Tier 2 cluster,
using the Armadillo C++ library. 

\appendix

\section{Adiabatic expansion of the currents}
\label{expansion appendix}

Here we write the explicit adiabatic expansion for the currents, in the background
\begin{equation}
 g = \frac{1}{\beta^2(r)}\[-\dd t^2 + \dd r^2 + \alpha^2(r)(\dd\theta^2 + \sin^2\theta \dd \phi^2)\]\,,
\end{equation} 
where the currents are defined by
\begin{align}
&j^\mu(x) = - \Tr[\gamma^\mu S(x, x')]\,,\\
& T^{\mu\nu}(x) = - \Tr\[\gamma^{(\mu}\, \ii D^{\nu)} S(x, x')\]\,,
\end{align}
with
\begin{equation}
 \sqrt{g}\,(\ii \gamma \cdot D - m) S(x, x') = \ii\, \delta(x - x')\,,
\end{equation} 
and $x= (t, r, \theta, \phi)$, $x'= (t \pm \ii s, r, \theta, \phi)$. Symmetrized over the sign of $s$.
{\small
\begin{equation}
\begin{split}
 \frac{\pi^2}{\beta^4} J^t = & -\frac{2 \Phi}{s^2} + L \(\frac{\alpha ' \Phi '}{3 \alpha }+\frac{\Phi ''}{6}\)-\\
 &-\frac{\Phi  \alpha ''}{6 \alpha }-\frac{\alpha ' \Phi '}{6 \alpha }-\frac{\Phi 
   \alpha '^2}{12 \alpha ^2}+ \frac{\Phi }{12 \alpha ^2}-\frac{\beta ' \Phi '}{6
   \beta }-\frac{\Phi ''}{12}-\frac{1}{3} \Phi ^3 + \mathcal{O}(s^2)
\end{split}
\end{equation} 
\begin{equation}
\begin{split}
 \frac{\pi^2}{\beta^4} T^t{}_{t} = & \frac{6}{s^4} + \frac{1}{s^2}\(\frac{\alpha ''}{6 \alpha }+\frac{\alpha '^2}{12 \alpha ^2}-\frac{1}{12 \alpha ^2}+3
   \Phi ^2\) +\\
&+L\(\frac{\alpha ^{(4)}}{120 \alpha }-\frac{\alpha ''^2}{240 \alpha ^2}+\frac{\alpha
   '^4}{240 \alpha ^4}+\frac{\alpha ^{(3)} \alpha '}{120 \alpha ^2}-\frac{\alpha
   '^2 \alpha ''}{60 \alpha ^3}-\frac{1}{240 \alpha ^4}+\frac{1}{12} \Phi '^2\) + \\
&+ \frac{\alpha ^{(4)}}{120 \alpha }-\frac{\alpha ^{(3)} \beta '}{60 \alpha  \beta
   }-\frac{\alpha '' \beta ''}{40 \alpha  \beta }+\frac{17 \alpha '' \beta
   '^2}{720 \alpha  \beta ^2}+\frac{\Phi ^2 \alpha ''}{12 \alpha }-\frac{\alpha
   ''^2}{240 \alpha ^2}-\frac{\beta ^{(3)} \alpha '}{60 \alpha  \beta
   }-\\& - \frac{\alpha '^2 \beta ''}{144 \alpha ^2 \beta }-\frac{11 \alpha ' \beta
   '^3}{360 \alpha  \beta ^3}+\frac{3 \alpha '^2 \beta '^2}{160 \alpha ^2 \beta
   ^2}+\frac{17 \alpha ' \beta ' \beta ''}{360 \alpha  \beta ^2}+\frac{\Phi 
   \alpha ' \Phi '}{3 \alpha }+\frac{\Phi ^2 \alpha '^2}{24 \alpha ^2}+\\&+\frac{\alpha
   '^4}{240 \alpha ^4}+\frac{\alpha ^{(3)} \alpha '}{120 \alpha ^2}-\frac{\alpha '
   \alpha '' \beta '}{72 \alpha ^2 \beta }-\frac{\alpha '^2 \alpha ''}{60 \alpha
   ^3}+\frac{\beta ''}{144 \alpha ^2 \beta }-\frac{\beta '^2}{288 \alpha ^2 \beta
   ^2}-\frac{\Phi ^2}{24 \alpha ^2}-\\&-\frac{1}{240 \alpha ^4}-\frac{\beta ^{(4)}}{240
   \beta }+\frac{7 \beta ''^2}{480 \beta ^2}+\frac{11 \beta '^4}{960 \beta
   ^4}+\frac{\beta ^{(3)} \beta '}{80 \beta ^2}-\frac{23 \beta '^2 \beta ''}{720
   \beta ^3}+\\&+\frac{1}{6} \Phi  \Phi ''+\frac{1}{24} \Phi '^2+\frac{\Phi ^4}{4} + \mathcal{O}(s^2)
\end{split}
\end{equation} 
\begin{equation}
\begin{split}
 \frac{\pi^2}{\beta^4} T^r{}_{r} &= -\frac{2}{s^4} + \frac{1}{s^2} \(-\frac{\alpha '^2}{12 \alpha ^2}+\frac{1}{12 \alpha ^2}-\Phi ^2\) + \\
 &+L\(\frac{\alpha ''^2}{240 \alpha ^2}+\frac{\alpha '^4}{240 \alpha ^4}-\frac{\alpha ^{(3)}
   \alpha '}{120 \alpha ^2}-\frac{1}{240 \alpha ^4}+\frac{1}{12} \Phi '^2\) - \\
 & -\frac{\alpha ^{(3)} \beta '}{120 \alpha  \beta }+\frac{\alpha '' \beta ''}{120
   \alpha  \beta }-\frac{\beta ^{(3)} \alpha '}{120 \alpha  \beta }-\frac{\alpha
   '^2 \beta ''}{60 \alpha ^2 \beta }-\frac{\alpha ' \beta '^3}{72 \alpha  \beta
   ^3}-\frac{7 \alpha '^2 \beta '^2}{480 \alpha ^2 \beta ^2}+\frac{\alpha '^3 \beta
   '}{60 \alpha ^3 \beta }+\\&+\frac{\alpha ' \beta ' \beta ''}{24 \alpha  \beta
   ^2}-\frac{\Phi  \alpha ' \Phi '}{6 \alpha }-\frac{\Phi ^2 \alpha '^2}{24
   \alpha ^2}+\frac{\alpha ' \alpha '' \beta '}{120 \alpha ^2 \beta }-\frac{\beta
   '^2}{288 \alpha ^2 \beta ^2}+\frac{\Phi ^2}{24 \alpha ^2}-\\&-\frac{\beta ''^2}{160
   \beta ^2}+\frac{11 \beta '^4}{960 \beta ^4}+\frac{\beta ^{(3)} \beta '}{80 \beta
   ^2}-\frac{\beta '^2 \beta ''}{40 \beta ^3}-\frac{1}{24} \Phi '^2-\frac{1}{12} \Phi
   ^4 + \mathcal{O}(s^2)
\end{split}
\end{equation} 
\begin{equation}
\begin{split}
 \frac{\pi^2}{\beta^4} T^\theta{}_{\theta} = \frac{\pi^2}{\beta^4} T^\phi{}_{\phi}&= -\frac{2}{s^4} + \frac{1}{s^2}\(-\frac{\alpha ''}{12 \alpha }-\Phi ^2\)+ \\ 
 & + L \(-\frac{\alpha '^4}{240 \alpha ^4}+\frac{\alpha '^2 \alpha ''}{120 \alpha
   ^3}+\frac{1}{240 \alpha ^4}-\frac{1}{12} \Phi '^2\) +\\
 &+ \frac{\alpha ^{(3)} \beta '}{120 \alpha  \beta }-\frac{\alpha '' \beta ''}{120
   \alpha  \beta }-\frac{7 \alpha '' \beta '^2}{1440 \alpha  \beta ^2}-\frac{\Phi
   ^2 \alpha ''}{24 \alpha }-\frac{\beta ^{(3)} \alpha '}{80 \alpha  \beta
   }+\frac{\alpha '^2 \beta ''}{120 \alpha ^2 \beta }+\\&+\frac{\alpha ' \beta
   '^3}{720 \alpha  \beta ^3}-\frac{\alpha '^2 \beta '^2}{120 \alpha ^2 \beta
   ^2}-\frac{\alpha '^3 \beta '}{120 \alpha ^3 \beta }+\frac{\alpha ' \beta '
   \beta ''}{90 \alpha  \beta ^2}-\frac{\Phi  \alpha ' \Phi '}{12 \alpha
   }+\\&+\frac{\alpha ' \alpha '' \beta '}{60 \alpha ^2 \beta }-\frac{\beta
   ^{(4)}}{240 \beta }+\frac{7 \beta ''^2}{480 \beta ^2}+\frac{11 \beta '^4}{960 \beta
   ^4}+\frac{\beta ^{(3)} \beta '}{80 \beta ^2}-\frac{23 \beta '^2 \beta ''}{720
   \beta ^3}-\\&-\frac{1}{12} \Phi  \Phi ''-\frac{1}{24} \Phi '^2-\frac{1}{12} \Phi ^4 + \mathcal{O}(s^2)
\end{split}
\end{equation} 
}

\section{How not to construct a gravitating quantum electron star}
\label{sec:cautionary}

In the course of developing the method of solution we described, we encountered many approaches that seemed natural choices, but revealed themselves to be complete blunders. We briefly describe them here
as a warning to any person that would get involved in this kind of problem in the future.

\subsection{Other regulators}
\label{other regulators}
We began our investigations looking at the problem with a frozen metric \cite{quantum_electron_star}. In that case, only the charge density is needed.
The charge density is a mildly divergent quantity, and hence not very sensitive to the regularization and renormalization procedure. 

In our preliminary work, we discretized the Dirac Hamiltonian using finite differences, and we considered a planar instead of spherical boundary. To avoid dealing with a continuous spectrum, we terminated the geometry with an artificial hard wall, following \cite{Sachdev:2011ze}. In this case the partial wave number $\ell$ is replaced by the transverse momentum $k$, and the radius $R$ of the sphere is replaced by the distance of the wall from the boundary.

In this setup, the lattice spacing $a$ provides a natural cutoff on the high frequency modes. The contribution of each $k$-mode to the charge density is finite in the limit $a \to 0$, because positive frequency and negative frequency modes make contributions with the opposite sign. The sum over the $k$ modes is logarithmically divergent, but it can easily be regulated with a hard cutoff on the momentum $k$. A change in this cutoff is equivalent to charge renormalization.

For the charge density this regularization and renormalization scheme works just fine, but it is not recommendable to use it when the geometry becomes dynamical. First of all, a hard wall termination of the geometry makes little sense when Einstein's equations are involved, so a geometric infrared regulator is needed. That is why we introduced the spherical geometry.

Second, the contribution of each $k$ mode to the energy density is not finite in the limit $a \to 0$, because positive frequency and negative frequency modes make contributions with the same sign. One needs to carry out some kind of subtraction to get rid of this infinity. But it is not obvious how to determine the counterterm. Since the infinity is strongly tied to the lattice physics, there is no procedure analogous to the adiabatic expansion that can give analytic information about the divergences. Moreover, even if one were able to obtain a finite subtracted quantity, it is not clear whether it would be a meaningful quantity, i.e. whether the subtraction procedure succeeded in restoring general covariance.

Understood the importance of general covariance as a guidance for the regularization and renormalization procedure, it is tempting to use a covariant regulator. For example, one can try the heat kernel regulator
\begin{align}
&J^\mu_0 = \expval{\bar \psi \gamma^\mu e^{-s^2 D\!\!\!\!/^2}\psi}\\
&T^{\mu\nu}_0 = \expval{\bar \psi \gamma^{(\mu}\ii D^{\nu)} e^{-s^2 D\!\!\!\!/^2} \psi}\,.
\end{align}

When using this regulator, all divergences are proportional to local geometric objects. The renormalization procedure consists in simply subtracting them, and general covariance is preserved throughout the process. In practice, one would expand over eigenfunctions of $\ovslash D$:
\begin{equation}
 \ovslash{D} \psi_n(x) = \lambda_n \psi_n(x)
\end{equation} 
and write
\begin{align}
&J^\mu_0 = \sum_n \bar\psi_n \gamma^\mu \psi_n e^{-s^2 \lambda^2}\\
&T^{\mu\nu}_0 = \sum_n \bar\psi_n \gamma^{(\mu}\ii D^{\nu)} \psi_n e^{-s^2 \lambda^2}\,.
\end{align}

Unfortunately, this approach has several problems. First of all, the operator $\ovslash{D}$ is not self adjoint, when the real time component $\Phi$ of the gauge field is non-zero. Consequently, the numerical diagonalization of $\ovslash{D}$ is problematic. Second, the label $n$ stands for the momenta in the time, radial and transverse directions. It is not possible to carry out the sum over any of these momenta analytically, even though there is time translational invariance and translational or spherical symmetry along the transverse directions. Therefore, the sum over $n$ truly is at best a double sum, with each term involving the diagonalization of a non-hermitian matrix, and a summation over the eigenfunctions. Moreover, the momentum in the radial direction is countinuous, because the operator $\ovslash{D}$ is non-compact at the boundary of AdS, so it is necessary to introduce a hard wall infrared regulator near the boundary. It is apparent that this is not quite the way to go.

A more promising approach is to use a Pauli-Villars regulator. One introduces a number of additional fictitious spinor fields, with appropriately chosen masses $M_i$ and statistics $\sigma_i$ (bosonic spinor fields may be needed), so that their contribution to the currents exactly cancels the contribution of the physical field at large energy. Explicitly:
\begin{align}
 &j^{\mu}_0(x) = \Tr\[\gamma^{\mu} S_m(x, x)\] + \sum_i \sigma_i \Tr\[\gamma^{\mu} S_{M_i}(x, x)\]\\
 &T^{\mu\nu}_0(x) = \Tr\[\gamma^{(\mu}\ii D^{\nu)} S_m(x, x)\] + \sum_i \sigma_i \Tr\[\gamma^{(\mu}\ii D^{\nu)} S_{M_i}(x, x)\]
\end{align} 

The masses and statistics can be found by studying the problem in flat space. In that case one has, for the stress tensor
\begin{equation}
 T^{\mu\nu}_0(x) = \int \dd^d p\ p^{\mu} p^{\nu}\( \frac{1}{p^2 + m^2} + \sum_i \sigma_i  \frac{1}{p^2 + M_i^2}\)\,.
\end{equation}
With an appropriate choice of $\sigma_i$ and $M_i$, the integral can be made convergent. For example, in two dimensions one can take
\begin{align}
 \sigma_1 = \sigma_2 = -1,\ \ \sigma_3 = 1, && M_1 = M_2 = M,\ \ M_3 = \sqrt{2M^2 - m^2}
\end{align}
Clearly the currents diverge in the large $M$ limit, but the coefficients of the divergent terms in a series expansion are local geometric objects, because this regulator is manifestly covariant. These terms can be subtracted, yielding well defined renormalized currents.

Pauli-Villars regulator makes it possible to express the currents in terms of the Dirac Hamiltonian. Introducing a set of eigenfunctions of the Dirac hamitonian
\begin{equation}
 H_{\ell, M_i} \psi_{n\ell i} = \omega_{n\ell i} \psi_{n\ell i}\,,
\end{equation} 
we have, for example,
\begin{equation}
 T^{tt}_0(x) = \frac{1}{2}\sum_{n,\ell} \sum_i \sigma_i \psi^\dag_{n\ell i}(x)\[\mathrm{abs}\, \omega_{n\ell i} -\Phi(x) \sign \omega_{n\ell i}\]\psi_{n\ell i}(x)\,,
\end{equation}
where we have included $m$ in the list of the masses $M_i$. The sum is convergent by construction, so the contribution of the higher frequency modes is less and less important. The problem has been reduced to a single sum, each term of which involves the diagonalization of a handful of hermitian matrices, and a summation over their eigenfunctions. This is a marked improvement over the heat kernel regulator. 

Unfortunately, the suppression of high frequency modes is only polynomial. This makes it necessary to compute a great number of terms in the $\ell$ and $n$ sum, and hence to diagonalize matrices of large size. Eventually, because of this reason, we choose to resort to point splitting regulation, which yields exponential suppression of the high energy modes. 

\subsection{Parallel transport}

The point-separated expressions
\begin{align}
&J^\mu_0(x) = \expval{\bar \psi(x') \gamma^\mu \psi(x)}\\
&T^{\mu\nu}_0(x) = \expval{\bar \psi(x') \gamma^{(\mu}\ii D^{\nu)} \psi(x)}
\end{align} 
may look awkward to the careful reader, because the spinors $\psi(x)$ and $\bar\psi(x')$ do not transform in a complementary way under gauge transformations and diffeomorphisms, and hence the bare currents are not tensors. One may be tempted to introduce a more covariant expression
\begin{align}\label{covariant currents}
&J^\mu_0(x) = \expval{\bar \psi(x') P(x', x) \gamma^\mu \psi(x)}\\
&T^{\mu\nu}_0(x) = \expval{\bar \psi(x') P(x', x) \gamma^{(\mu}\ii D^{\nu)} \psi(x)}\,,
\end{align} 
where $P$ is the spinor parallel transport, satisfying
\begin{equation}
\left\{
\begin{aligned}
 &D_\mu P(x, x') = 0\\
 &P(x, x) = 1\,.
\end{aligned}\right.
\end{equation} 

While there is certainly nothing wrong in doing so, it is not necessary, the reason being that the subtraction of the adiabatic expansion cancels all the covariance-breaking effects of the regulator. To show this explicitly, let us assume the covariant definitions (\ref{covariant currents}), and show that the parallel transport has no effect after subtraction of the adiabatic expansion. Let us consider the U$(1)$ current. We have
\begin{align}
J^\mu_0(x) = -\Tr\[\gamma^\mu S(x, x') P(x', x)\]\\
\end{align}
The propagator $S(x, x')$ diverges as $x'$ approaches $x$, and we have\footnote{There are also logarithmic divergences, which do not matter for the following.}
\begin{equation}
 S(x, x') = \frac{1}{s^3} S_3(x,x') + \frac{1}{s^2} S_2(x,x') + \ldots + S_0(x, x')\,,
\end{equation}
where $s^2 = (x - x')^\mu(x - x')_\mu$ and the $S_i$ have a finite limit as $s \to 0$. On the other hand $P(x, x) = 1$, so
\begin{equation}
 P(x, x') = 1 + s\ P_1(x, x') + s^2 P_2(x, x') + \ldots
\end{equation} 
Therefore, the portion of the current that depends on $P$ and that does not vanish as $s \to 0$ is
\begin{equation}
 J^\mu_0(x)|_P = - \frac{1}{s^2} \Tr[S_3 P_1] - \frac{1}{s} \(\Tr[S_3 P_2] + \Tr[S_2 P_1]\) - \(\Tr[S_3 P_3] + \Tr[S_2 P_2] + \Tr[S_1 P_1]\)\,.
\end{equation} 
This expression depends only on the divergent terms of $S$, which are captured in full by the adiabatic expansion. Therefore, after subtraction of the adiabatic expansion and the limit $s \to 0$, there is no dependence of $P$ left, and the result is the same as if it had not been included from the beginning.

Besides complicating the algebra unnecessarily, inclusion of the parallel transport has another undesirable consequence. As shown in section (\ref{conservation section}), if the point splitting is in the time direction, with constant coordinate separation, the bare stress tensor is covariantly conserved. This property is lost if the parallel transport is included. Obviously it is restored by the regularization and renormalization procedure, but there is some advantage in having it throughout the process.

 \subsection{WKB instead of adiabatic expansion}
 
 It is tempting to try to use the WKB approximation to determine the high energy behavior of the eigenfunctions of the Dirac Hamiltonian. If that were possible, one could subtract the high energy behavior directly in the mode sum, for example
 \begin{equation}
 J^t  = \frac{1}{e_t e_r e_s^2} \sum_{n \ell} \frac{|\ell|}{2\pi}\  \[\psi^\dag_{n\ell}(r)\psi_{n\ell}(r) - \psi^{WKB\dag}_{n\ell}(r)\psi^{WKB}_{n\ell}(r)\]\,,\\ 
 \end{equation} 
 and compute the renormalized currents directly as an altogether finite sum, without the need for any other subtraction or limit, provided that the subtraction can be shown to preserve general covariance. This program works when the background is spatially uniform, and depends on time, but it fails when there is non-trivial spatial dependence. 
 
The issue can be demonstrated 
(by replacing the Dirac equation) with
the more familiar Schr\"odinger equation, 
in more than one dimension.
It arises even if the Schr\"odinger operator (\ie\ the potential) depends 
only on one variable $r$.  
So consider
\begin{equation}\label{eq:schrod} - \nabla^2 \psi(r, y) + \( V(r) - E \)  \psi(r, y) = 0 ~.\end{equation}
Translation invariance in $y$ -- the proxy for the QFT spatial slices,
of which there can be more than one for the present discussion -- 
and linearity of \eqref{eq:schrod} allow us to Fourier decompose: 
\begin{equation} \psi(r,y) = e^{ \ii k y } \psi(r) \end{equation}
so that 
\begin{equation}\label{eq:schrodk} - \psi''(r) + \( V(r) + k^2  - E \)  \psi(r) = 0 ~.\end{equation}
For $E \gg V(x) + k^2 $ we may use a WKB ansatz: 
\begin{equation} \psi(r) = { \mathcal{N} \over \sqrt{q(x) } } \exp{\( \ii \int_{-\infty}^r q(r') \dd r' \) } ~, \end{equation}
and $q(r)$ must satisfy
\def\eps{\epsilon}s
\begin{equation} 
q^2(r) - \( E - k^2 - V(r) \) -  \eps \( { 3\over 4} { q'(r)^2 \over q(r)^2} - {q''(r)\over 2 q(r)}  \) ~. 
\end{equation}
with $\eps = 1$ is a book-keeping parameter.
The WKB approximation treats the last two terms as an approximation by writing
\begin{equation} q(r) = q_0(r) + \eps q_1(r) + \mathcal{O}(\eps^) ~. \end{equation}
Solving order-by-order in $\eps$ gives
\begin{eqnarray}
q_0(r) &=& \sqrt{ E - k^2 - V(r) } \cr \cr
q_1(r) & =&  { 1\over 8} { V''(r) \over q_0(r)^3} + {1\over 32} {V'(r)^2 \over q_0(r)^5} ~.  
\end{eqnarray}
In this approximation, the contribution to the particle density 
from a given mode is 
\begin{equation} \norm{\psi(r, y)}^2 = { \mathcal{N}\over q(r) } 
= { \mathcal{N}\over q_0(r) } - \eps { \mathcal{N} q_1(r) \over q_0^2 }  |_{\eps=1}
= \mathcal{N} \( {1\over q_0(r) } - { 1\over 8 } { V''(r) \over q_0(r)^5} 
+ { 1\over 32} { V'(r)^2 \over q_0(r)^7} \) . \end{equation}

Now assume for argument that $V$ and $V'$ vanish at $r \to \pm \infty$, 
so that the states at large $r$ are plane waves, which we can 
take to be incoming: 
\begin{equation} p = q(- \infty) = \sqrt{ E - k^2 } .\end{equation}
One can verify by putting the system in a box that the correct
integration measure (for adding up the contributions of the modes
to the total density or total energy) is 
\begin{equation} \int \dbar k ~\dbar p \equiv \int { \dd k \over 2\pi} { \dd p \over 2\pi}  ,\end{equation}
so we can label the states by $(E,k)$ rather than $(p,k)$.
So
\begin{equation} q_0(r) = \sqrt{p^2 - V(r)} \end{equation}
and, for example, the (heat-kernel) regulated energy density is
\begin{equation} 
\varepsilon(r,y) = \int \dbar k \dbar p E(k, p, r) \norm{\psi(r,y)}^2 e^{ - s E(k,p,r)} 
\end{equation}
with 
\begin{equation} E(k,p,r) \equiv \sqrt{ k^2 + p^2 + V(r)} \end{equation}
and 
\begin{equation} \norm{\psi(r, y)}^2 
= \mathcal{N} \( {1\over \sqrt{p^2 - V(r)}  } - { 1\over 8 } { V''(r) \over \(p^2 - V(r)\)^{5/2} }
+ { 1\over 32} { V'(r)^2 \over\(p^2 - V(r)\) ^{5/2}} \) . \end{equation}
This expression has a non-integrable singularity when $p^2 \to V(r)$.  
This is to be expected, since the WKB approximation is valid in the limit $ p^2 \gg V(r)$.

The problem for our purposes, is that we need to integrate over $p$.
We could consider restricting the integral to large $ |p|> M $.
But this excludes a slice of the integration region
{\it that includes very high energy modes}, 
and hence modifies the $s$-dependence of the integral.
The problem is that the singularity from the breakdown of WKB 
is present not only for small $k^2+p^2$, and one cannot 
exclude it without excluding high-energy modes.

The point, then, is that no matter how large $E$ is, 
there are still modes of large enough $k$ to cause the WKB 
approximation to break down.

One can attempt a higher-dimensional generalization of WKB. 
Unfortunately, this does not work because WKB in higher dimensions
(at least as we understand it) is non-local: 
Suppose we want to solve
\begin{equation} 
\label{eq:schrod2} - \nabla^2 \psi(r, y) + \( V(r,y) - E \)  \psi(r, y) = 0 ~\end{equation}
and substitute 
\begin{equation}\label{eq:wkb2} \psi(r,y) = A(r,y) \exp{\( \ii \int_\infty^{(r,y)} \vec q(\vec r) \cdot \dd \vec r \) } ~.\end{equation}
We arrive at the system of equations
\begin{equation}
\begin{cases}
{ 1\over A} \nabla^2 A + \( E - V + \vec q^2 \)  =   0 \cr 
\vec \nabla \cdot \( A^2 \vec q \) = 0 \cr
\partial_r q_y - \partial_y q_r= 0 
\end{cases}.
\end{equation}
where the last equation ensures that the phase in \eqref{eq:wkb2} does
not depend on the path.
The amplitude $A(r,y)$, however, depends on the potential $V(r',y')$ 
at arbitrarily distant points.

We note in passing that 
this problem seems to present an obstruction
to a systematic implementation of 
the expansion studied in \cite{Medvedyeva:2013rpa}.

\subsection{Localization of eigenstates of the lattice Dirac Hamiltonian}

\begin{figure}[h]
\begin{center}
\includegraphics[height=100pt]{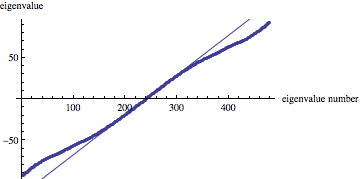} 
\includegraphics[height=100pt]{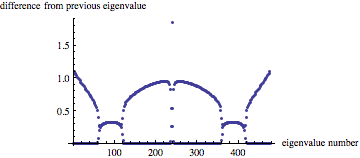} 
\includegraphics[height=100pt]{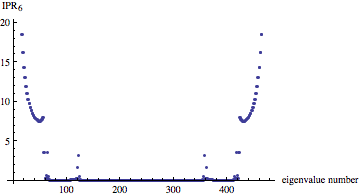}
\caption{\label{fig:iprs}
In this example, $f_t = 1 + .3 \cos z + .2 \cos 2z, f_x = 1 + .4 \cos z - .2 \cos 2z$.  $n = 249$ sites.
In the first plot, the line represents the continuum dispersion.
The second plot shows successive differences of eigenvalues; 
the zeros in this plot indicate eigenvalues that come in pairs 
related by parity, as is true of all modes in the continuum.
Note the presence of several bands of parity-paired states.
}
\end{center}
\end{figure}

In this section we explain why it is necessary to 
use the point-splitting regulator $s$ 
in addition to the lattice regularization
of the Dirac operator.
One might consider simply using a lattice regularization of the Dirac Hamiltonian,
with near-neighbor derivatives, or some improvement thereof, 
for example using spectral methods.
However, only a fraction of the eigenstates of the resulting lattice operator
have anything at all to do with the continuum limit.
This fact is visible in a plot of the eigenvalues versus mode number (Fig.~\ref{fig:iprs}a)
and of the successive differences between eigenvalues (Fig.~\ref{fig:iprs}b),
for a 1+1 dimensional case.
In the middle of the spectrum, the dispersion is linear and agrees with the expected
slope in the continuum (straight line).  Further, the eigenstates come in pairs 
consisting of a left-mover and a right-mover, as one expects from plane waves in the continuum.  
Away from the middle of the 
spectrum, we find only 1d representations of the parity operator.
We observe that there is a sharp boundary where the degeneracy ends.

That this is a sharp `mobility edge' in the spectrum, 
separating extended and localized states, can be seen as follows.
A measure of localization is the Inverse Participation Ratio (IPR), defined as 
$$ \text{IPR}_k[\psi] = \int dx ~| \psi(x)|^{2k} $$
for a normalized wavefunction $ \int  dx ~| \psi(x)|^{2} = 1. $
The results can be seen in Fig.~\ref{fig:iprs}.

Leaving out the localized modes would make it impossible to resolve the identity,
and one cannot construct smooth sources without them.
The problem of central interest to us is the 4d bulk Dirac operator, 
with translation invariance in the QFT spatial directions.  
Each momentum mode satisfies a 1+1 Dirac equation (with a complex $k$-dependent mass).
Therefore, this same problem persists for us in higher dimensions.


%
\begin{figure}[h]
\begin{center}
\includegraphics[height=150pt]{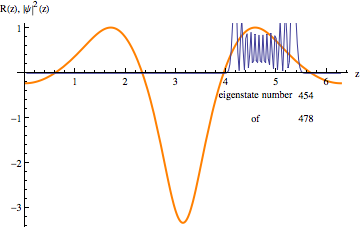} 
\caption{\label{fig:curvature}
\small
A plot of the curvature (orange, thick) in the example above,
along with one of the visibly localized eigenfunctions (blue, thin).
}
\end{center}
\end{figure}


More explicitly, 
we consider a massless Dirac field in the geometry
$$ds^2 = f_t(z) dt^2 + f_z(z) dz^2 + f_x(z) d\vec x^2 $$ 
with electrostatic potential $\Phi$.
It will be useful to write veilbeins $ e_\mu^2 \equiv f_\mu$.
(Below we will take $d=0$,  $ z \simeq z + 2\pi$ is a circle
and $f_t, f_z$ are chosen to be periodic, and $\Phi=0$.)
In terms of 
$$ \Psi(z, \vec x, t) = e^{ -i \omega t + i \vec k \cdot \vec x } \sqrt{ e_t \over \prod_{\mu=t,z,\vec x} e_\mu} \psi(z) $$
the Dirac equation $( \Dslash + m ) \Psi = 0 $ can be written as 
$ H \psi = \omega \psi$.
In this basis, the Dirac Hamiltonian (at fixed $k$) is 
self-adjoint with respect to the usual inner product
$$ \( \psi_1, \psi_2 \) = \int dz ~\psi_1^\star(z) \psi_2(z) .$$

It is convenient to study a different basis, where derivatives of the metric do not appear: 
$$ \Psi(z, \vec x, t) = e^{ -i \omega t + i \vec k \cdot \vec x } \sqrt{ e_z \over \prod_{\mu=t,z,\vec x} e_\mu} 
\tilde \psi(z) $$
in which 
the Dirac equation $( \Dslash + m ) \Psi = 0 $ can be written as 
$ \tilde H \tilde \psi = \omega \tilde \psi$,
where $\tilde H$ has no derivatives of $e_\mu$.
We have
$$ \tilde \psi = \sqrt{ e_t \over e_z} \psi $$
and therefore
$$ \tilde H \sqrt{ e_t \over e_z} \psi  = \omega \sqrt{ e_t \over e_z} \psi $$
which means
\begin{eqnarray} H 
=\sqrt{ e_z \over e_t} \tilde H \sqrt{ e_t \over e_z} 
=
\underbrace{\sqrt{ e_t \over e_z}}_W \underbrace{ {e_z \over e_t}  \tilde H}_{\hat H} \underbrace{\sqrt{ e_t \over e_z}}_{W}
\end{eqnarray}
Here $W$ is diagonal in position space,
and $\hat H$ is a Hermitean matrix, free from derivatives of $e_\mu$,
with which $H$ is isospectral.
Explicitly, with $k=0$: 
$$ 
H = 
\begin{pmatrix}
- m \sqrt{f_t} + \Phi & - \({f_t\over f_z}\)^{1/4} \partial_z \({f_t\over f_z}\)^{1/4}
\cr
\({f_t\over f_z}\)^{1/4} \partial_z \({f_t\over f_z}\)^{1/4}
&  m \sqrt{f_t} + \Phi
\end{pmatrix}
= W  \begin{pmatrix}
- m \sqrt{f_z} + \Phi \sqrt{f_z\over f_t} 
& - \partial_z 
\cr
 \partial_z &  m \sqrt{f_z} + \Phi \sqrt{f_z\over f_t} 
\end{pmatrix}
.
$$

A straightforward way to put this operator on the lattice is simply to replace
$ \psi(z) $ with a column vector of values at equidistant points $\psi(z_i)$, 
and to replace 
$$
 \begin{pmatrix}
\cdot
& - \partial_z 
\cr
 \partial_z & \cdot 
\end{pmatrix}
\mapsto 
 \begin{pmatrix}
\cdot
& B 
\cr
 B^t & \cdot 
\end{pmatrix}
$$
with
\begin{equation}
 B = \frac{1}{\Delta z}
\begin{pmatrix}
  -1 &  1 &  0 &  0 &  0 \\
   0 & -1 &  1 &  0 &  0 \\
   0 &  0 & -1 &  1 &  0 \\
   0 &  0 &  0 & -1 &  1 \\
   0 &  0 &  0 &  0 & -1 \\
\end{pmatrix} 
\end{equation}

\begin{singlespace}
\bibliographystyle{ssg}
\bibliography{GES}
\end{singlespace}

\end{document}